\documentclass[12pt]{article}
\pdfoutput=1
\usepackage[colorlinks,linkcolor=Blue,citecolor=Blue,bookmarks,bookmarksnumbered]{hyperref}
\usepackage[scaled=0.85]{helvet}
\usepackage{amsmath,amssymb,accents,mathrsfs,XoohmE}
\usepackage{graphicx,color}
\graphicspath{{Figures/}}
\usepackage{booktabs}
\usepackage{multirow}
\usepackage{placeins}
\usepackage{amsmath}
\usepackage{subfigure}

\usepackage{XoohmE}

\definecolor{Green}  {rgb}{0.10,0.70,0.10} 
\definecolor{Orange} {rgb}{1.00,0.50,0.15} 
\definecolor{Red}    {rgb}{0.90,0.00,0.12} 
\definecolor{Purple} {rgb}{0.50,0.25,0.55} 
\definecolor{Turque} {rgb}{0.00,0.65,0.85} 
\definecolor{Blue}   {rgb}{0.00,0.00,1.00} 
\definecolor{Magenta}{rgb}{1.00,0.00,1.00} 
\definecolor{Gold}   {rgb}{1.00,0.75,0.25} 
\definecolor{Seaweed}{rgb}{0.01,0.24,0.09} 
\definecolor{Brown}  {rgb}{0.43,0.26,0.32} 
\definecolor{grey1}  {rgb}{0.20,0.20,0.20} 
\definecolor{grey2}  {rgb}{0.40,0.40,0.40} 
\definecolor{grey3}  {rgb}{0.60,0.60,0.60} 
\definecolor{grey4}  {rgb}{0.80,0.80,0.80} 
\definecolor{grey5}  {rgb}{0.90,0.90,0.90} 
\def\C#1#2{{\ifcase#1\or
             \color{Green}\or \color{Orange}\or \color{Red}\or
              \color{Purple}\or \color{Turque}\or \color{Blue}\or
               \color{Magenta}\or \color{Gold}\or \color{Seaweed}\or
                \color{Brown}\or\color{grey1}\or\color{grey2}\or
                 \color{grey3}\else\color{grey4}\fi#2}}

\definecolor{Slate} {rgb}{0.00,0.45,0.55}

\newcommand{\rdm}{\rd_{\min}}

\newcommand\GR{\mathcal{GR}}

\def\fracm#1#2{\hbox{\large{${\frac{{#1}}{{#2}}}$}}}

\def\be{\begin{equation}}
\def\ee{\end{equation}}
\newcommand{\bea}{\begin{eqnarray}}
\newcommand{\eea}{\end{eqnarray}}
\newcommand{\ena}{\end{eqnarray}}

\def\pp{{\mathchoice
          {
              \kern 1pt%
              \raise 1pt
              \vbox{\hrule width5pt height0.4pt depth0pt
                    \kern -2pt
                    \hbox{\kern 2.3pt
                          \vrule width0.4pt height6pt depth0pt
                          }
                    \kern -2pt
                    \hrule width5pt height0.4pt depth0pt}%
                    \kern 1pt
           }
            {
              \kern 1pt%
              \raise 1pt
              \vbox{\hrule width4.3pt height0.4pt depth0pt
                    \kern -1.8pt
                    \hbox{\kern 1.95pt
                          \vrule width0.4pt height5.4pt depth0pt
                          }
                    \kern -1.8pt
                    \hrule width4.3pt height0.4pt depth0pt}%
                    \kern 1pt
            }
            {
              \kern 0.5pt%
              \raise 1pt
              \vbox{\hrule width4.0pt height0.3pt depth0pt
                    \kern -1.9pt  
                    \hbox{\kern 1.85pt
                          \vrule width0.3pt height5.7pt depth0pt
                          }
                    \kern -1.9pt
                    \hrule width4.0pt height0.3pt depth0pt}%
                    \kern 0.5pt
            }
            {
              \kern 0.5pt%
              \raise 1pt
              \vbox{\hrule width3.6pt height0.3pt depth0pt
                    \kern -1.5pt
                    \hbox{\kern 1.65pt
                          \vrule width0.3pt height4.5pt depth0pt
                          }
                    \kern -1.5pt
                    \hrule width3.6pt height0.3pt depth0pt}%
                    \kern 0.5pt
            }
        }}

\def\mm{{\mathchoice
                       {
                             \kern 1pt
               \raise 1pt    \vbox{\hrule width5pt height0.4pt depth0pt
                                  \kern 2pt
                                  \hrule width5pt height0.4pt depth0pt}
                             \kern 1pt}
                       {
                            \kern 1pt
               \raise 1pt \vbox{\hrule width4.3pt height0.4pt depth0pt
                                  \kern 1.8pt
                                  \hrule width4.3pt height0.4pt depth0pt}
                             \kern 1pt}
                       {
                            \kern 0.5pt
               \raise 1pt
                            \vbox{\hrule width4.0pt height0.3pt depth0pt
                                  \kern 1.9pt
                                  \hrule width4.0pt height0.3pt depth0pt}
                            \kern 1pt}
                       {
                           \kern 0.5pt
             \raise 1pt  \vbox{\hrule width3.6pt height0.3pt depth0pt
                                  \kern 1.5pt
                                  \hrule width3.6pt height0.3pt depth0pt}
                           \kern 0.5pt}
                       }}

\def\ad{{\kern0.5pt
                   \alpha \kern-5.05pt \raise5.8pt\hbox{$\textstyle.$}\kern
0.5pt}}

\def\bd{{\kern0.5pt
                   \beta \kern-5.05pt \raise5.8pt\hbox{$\textstyle.$}\kern
0.5pt}}

\def\qd{{\kern0.5pt
                   q \kern-5.05pt \raise5.8pt\hbox{$\textstyle.$}\kern
0.5pt}}
\def\Dot#1{{\kern0.5pt
     {#1} \kern-5.05pt \raise5.8pt\hbox{$\textstyle.$}\kern
0.5pt}}

\catcode`@=11
\def\un#1{\relax\ifmmode\@@underline#1\else
        $\@@underline{\hbox{#1}}$\relax\fi}
\catcode`@=12

\def\a{\alpha}
\def\b{\beta}
\def\c{\chi}
\def\d{\delta}
\def\e{\epsilon}

\def\g{\gamma}

\def\i{\iota}
\def\j{\psi}
\def\k{\kappa}
\def\l{\lambda}
\def\m{\mu}
\def\n{\nu}
\def\o{\omega}

\def\r{\rho}
\def\s{\sigma}
\def\t{\tau}

\def\L{\Lambda}
\def\O{\Omega}
\def\P{\Pi}

\def\S{\Sigma}

\def\ca{{\cal A}}

\def\cf{{\cal F}}

\def\dslash{\not{\hbox{\kern-2pt $\partial$}}}
\def\Dslash{\not{\hbox{\kern-4pt $D$}}}
\def\pslash{\not{\hbox{\kern-2.3pt $p$}}}
 \newtoks\slashfraction
 \slashfraction={.13}
 \def\slash#1{\setbox0\hbox{$ #1 $}
 \setbox0\hbox to \the\slashfraction\wd0{\hss \box0}/\box0 }
 
\def\kcr{{\hbox{\ro \char'170}}}                
\def\ktl{{\hbox{\ro \char'170}}}        
\def\ktr{{\hbox{\ro \char'170}}}        
\def\kbl{{\hbox{\ro \char'170}}}        
\def\kbr{{\hbox{\ro \char'170}}}        

\def\plpl{\raise-2pt\hbox{$\raise3pt\hbox{$_+$}\hskip-6.67pt\raise0.0pt
\hbox{$^+$}\hskip 0.01pt$}}
\def\mimi{\raise-2pt\hbox{$\raise3pt\hbox{$_-$}\hskip-6.67pt\raise0.0pt
\hbox{$^-$}\hskip 0.01pt$}} 

\def\bo{{\raise.15ex\hbox{\large$\Box$}}}               
\def\pa{\partial}                                       
\def\TH{{\raise.2ex\hbox{$\displaystyle \bigodot$}\mskip-4.7mu \llap H \;}}
\def\face{{\raise.2ex\hbox{$\displaystyle \bigodot$}\mskip-2.2mu \llap {$\ddot
        \smile$}}}                                      

\def\dt#1{\on{\hbox{\bf .}}{#1}}                
\def\Dot#1{\dt{#1}}

\def\Tilde#1{\widetilde{#1}}                    
\def\Bar#1{\overline{#1}}                       
\def\leftrightarrowfill{$\mathsurround=0pt \mathord\leftarrow \mkern-6mu
        \cleaders\hbox{$\mkern-2mu \mathord- \mkern-2mu$}\hfill
        \mkern-6mu \mathord\rightarrow$}
\def\dvec#1{\vbox{\ialign{##\crcr
        \leftrightarrowfill\crcr\noalign{\kern-1pt\nointerlineskip}
        $\hfil\displaystyle{#1}\hfil$\crcr}}}           
\def\dt#1{{\buildrel {\hbox{\LARGE .}} \over {#1}}}     

\def\fracm#1#2{\hbox{\large{${\frac{{#1}}{{#2}}}$}}}
\def\sfrac#1#2{{\vphantom1\smash{\lower.5ex\hbox{\small$#1$}}\over
        \vphantom1\smash{\raise.4ex\hbox{\small$#2$}}}} 
\def\bfrac#1#2{{\vphantom1\smash{\lower.5ex\hbox{$#1$}}\over
        \vphantom1\smash{\raise.3ex\hbox{$#2$}}}}       
\def\afrac#1#2{{\vphantom1\smash{\lower.5ex\hbox{$#1$}}\over#2}}    

\def\pa{\partial}

\def\ad{{\dot{\alpha}}}
\def\bd{{\dot{\beta}}}

 \font\rOpe=cmsy10                        
 \def\ktl{{\hbox{\rOpe\char'170}}}        
 \def\kbl{{\hbox{\rOpe\char'170}}}        
 \def\kcr{{\reflectbox{\rOpe\char'170}}}        
 \def\ktr{{\reflectbox{\rOpe\char'170}}}        
 \def\kbr{{\reflectbox{\rOpe\char'170}}}        
 \def\Border{\vbox{\hsize0pt
        \setlength{\unitlength}{1mm}
        \newcount\xco
        \newcount\yco
        \xco=-21
        \yco=12
        \begin{picture}(0,0)(-7.5,0)
        \put(\xco,\yco){$\ktl$}
        \advance\yco by-1
        {\loop
        \put(\xco,\yco){$\kcr$}
        \advance\yco by-2
        \ifnum\yco>-240
        \repeat
        \put(\xco,\yco){$\kbl$}}
        \xco=170
        \yco=12
        \put(\xco,\yco){$\ktr$}
        \advance\yco by-1
        {\loop
        \put(\xco,\yco){$\kcr$}
        \advance\yco by-2
        \ifnum\yco>-240
        \repeat
        \put(\xco,\yco){$\kbr$}}
        \put(-19.5,13){\scalebox{.6065}{%
         University of Maryland Center for String and Particle  Theory \&\ Physics Department%
        |University of Maryland Center for String and Particle  Theory \&\ Physics Department}}
        \put(-19.5,-241.5){\scalebox{.5835}{%
         ****University of Maryland * Center for String and
         Particle  Theory* Physics Department****University of Maryland *Center
        for String and Particle  Theory* Physics Department}}
        \end{picture}
        \par\vskip-8mm}}
\definecolor{UMred}{rgb}{.9,.05,.2}
\definecolor{HUblue}{rgb}{.0,.3,.7}

\definecolor{Red}    {rgb}{0.90,0.00,0.12} 
\definecolor{Blue}   {rgb}{0.00,0.00,1.00} 
\definecolor{Green}  {rgb}{0.10,0.70,0.10} 
\definecolor{Turque} {rgb}{0.00,0.65,0.85} 
\definecolor{Orange} {rgb}{1.00,0.50,0.15} 
\definecolor{Magenta}{rgb}{1.00,0.00,1.00} 
\definecolor{Gold}   {rgb}{1.00,0.75,0.25} 
\definecolor{Seaweed}{rgb}{0.01,0.24,0.09} 
\definecolor{Purple} {rgb}{0.50,0.25,0.55} 
\definecolor{Brown}  {rgb}{0.43,0.26,0.32} 
\definecolor{grey1}  {rgb}{0.20,0.20,0.20} 
\definecolor{grey2}  {rgb}{0.40,0.40,0.40} 
\definecolor{grey3}  {rgb}{0.60,0.60,0.60} 
\definecolor{grey4}  {rgb}{0.80,0.80,0.80} 
\definecolor{grey5}  {rgb}{0.90,0.90,0.90} 
\def\C#1#2{{\ifcase#1\or
             \color{Red}\or \color{Green}\or \color{Blue}\or\
              \color{Turque}\or \color{Orange}\or \color{Magenta}\or 
               \color{Gold}\or \color{Seaweed}\or \color{Purple}\or
                \color{Brown}\or\color{grey1}\or\color{grey2}\or
                 \color{grey3}\else\color{grey4}\fi#2}}

\definecolor{Slate} {rgb}{0.00,0.45,0.55}

\newdimen\parshift\parshift=\parindent
\catcode`@=11
 \long\def\@footnotetext#1{\insert\footins{\reset@font\footnotesize
           \interlinepenalty\interfootnotelinepenalty\splittopskip%
            \footnotesep\splitmaxdepth\dp\strutbox\floatingpenalty\@MM%
             \hsize\columnwidth\addtolength{\hsize}{-2\parindent}
              \@parboxrestore\protected@edef\@currentlabel%
              {\csname p@footnote\endcsname\@thefnmark}%
                \color@begingroup%
                 \@makefntext{\rule\z@\footnotesep\ignorespaces#1%
                  \@finalstrut\strutbox}%
                \color@endgroup}}
 \long\def\@makefntext#1{\hglue\parshift%
           \vbox{\noindent\baselineskip=11pt plus.5pt minus.5pt\hb@xt@0em{\hss\@makefnmark\kern1pt}#1}}
\catcode`@=12

\newskip\humongous \humongous=0pt plus 1000pt minus 1000pt
\def\caja{\mathsurround=0pt}
\def\eqalign#1{\,\vcenter{\openup2\jot \caja
        \ialign{\strut \hfil$\displaystyle{##}$&$
        \displaystyle{{}##}$\hfil\crcr#1\crcr}}\,}
\newif\ifdtup

\makeatletter
\def\section{\@startsection{section}{1}{\z@}
        {3ex plus-1ex minus-.2ex}{1pt plus1pt}{\large\sf\bfseries\boldmath}}
\def\subsection{\@startsection{subsection}{2}{\z@}
         {1.5ex plus-1ex minus-.2ex}{0.01pt plus1pt}{\sf\slshape}}
\def\subsubsection{\@startsection{subsubsection}{3}{\z@}
          {1.5ex plus-1ex minus-.2ex}{0.01pt plus0.2pt}{\sf\boldmath}}
\def\paragraph{\@startsection{paragraph}{4}{\z@}
           {.75ex \@plus.5ex \@minus.2ex}{-2mm}{\sf\bfseries\boldmath}}
\makeatother

 \allowdisplaybreaks
 \seceq

\usepackage{lipsum}
\usepackage{listings}
\definecolor{MyDarkGreen}{rgb}{0.0,0.4,0.0} 
\newcommand{\perlscript}[2]{
\begin{itemize}
\item[]\lstinputlisting[caption=#2,label=#1]{#1.pl}
\end{itemize}
}

\begin{document}

\thispagestyle{empty}
\noindent{\small
\hfill{HET-1788  \\ 
$~~~~~~~~~~~~~~~~~~~~~~~~~~~~~~~~~~~~~~~~~~~~~~~~~~~~~~~~~~~~~~~~~$
$~~~~~~~~~~~~~~~~~~~~~~~~~~~~~~~~~~~~~~~~~~~~~~~~~~~~~~~~~~~~~~~~~$
{}
}
\vspace*{8mm}
\begin{center}
{\large \bf
Adinkra Height Yielding Matrix Numbers:   \\[2pt]
Eigenvalue Equivalence Classes   \\[2pt]
for Minimal Four-Color Adinkras   \\[2pt]
}   \vskip1in
{\large {
S.\ James Gates, Jr.\footnote{sylvester${}_-$gates@brown.edu}$^{a}$},
Yangrui Hu\footnote{yangrui\_hu@brown.edu}$^{a}$, and Kory Stiffler\footnote{kory\_stiffler@brown.edu}$^{a}$
}
\\*[12mm]
\emph{
\centering
$^{a}$Department of Physics, Brown University,
\\[1pt]
Box 1843, 182 Hope Street, Barus \& Holley 545,
Providence, RI 02912, USA 
}
 \\*[40mm]
{ ABSTRACT}\\[4mm]
\parbox{142mm}{\parindent=2pc\indent\baselineskip=14pt plus1pt
An adinkra is a graph-theoretic representation of spacetime supersymmetry. 
Minimal four-color valise adinkras have been extensively studied due to their 
relations to minimal 4D, $\cal N$ = 1 supermultiplets.  Valise adinkras, although
an important subclass, do not encode all the information present when a 4D 
supermultiplet is reduced to 1D.  Eigenvalue equivalence classes for valise 
adinkra matrices exist, known as $\chi_{\rm o}$ equivalence classes, where 
valise adinkras within the same $\chi_{\rm o}$ equivalence class are isomorphic 
in the sense that adinkras within a $\chi_{\rm o}$-equivalence class can be 
transformed into each other via field redefinitions of the nodes.  We extend 
this to non-valise adinkras, via Python code, providing a complete eigenvalue 
classification of ``node-lifting'' for all 36,864 valise adinkras associated with 
the Coxeter group $BC{}_4$. We term the eigenvalues associated with these 
node-lifted adinkras Height Yielding Matrix Numbers (HYMNs) and introduce 
HYMN equivalence classes. These findings have been summarized in a 
\emph{Mathematica} notebook that can found at the HEPTHools 
\href{https://hepthools.github.io/Data/}{Data Repository} on GitHub.
}
 \end{center}
\vfill
\vfill
\clearpage
\section{Introduction}

The discovery of the role played by the ${\cal GR}$(d,$N$) ``Garden Algebras'' 
\cite{GRana1,GRana2} in the generation of representations of spacetime SUSY
triggered the introduction of adinkras \cite{Adnk1} in order to allow the technology 
of graph theory to include and expand prior existing results.  Among these results 
was an algorithm for generating dimensions of the minimal linear representations 
for all possible supersymmetrical systems.  This is encoded in a function denoted 
by d${}_{min}(N)$ \cite{GRana1,GRana2} where $N$ is the number of one 
dimensional supercharges.  In an adinkra graph, links denote the orbits of the 
one-dimensional projections of space-time functions in a supermultiplet under 
the action of space-time supercharges.  The nodes of an adinkra graph are 
projections of the fields of a supermultiplet when their functional dependence 
is restricted to solely depend on time.  Since their introduction, adinkras have 
also opened gateways to insights in  the physics and mathematics of SUSY.

{\it {
On the physics side, one of the most striking implications uncovered is every 
off-shell linear supersymmetrical system that allows for the solution of the initial 
value problem involves adinkras containing error-correcting codes
\cite{codes1,codes2,codes3}. This apparently implies an information-theoretic 
foundation is present in all supersymmetrical particle, field, and string theories.}}

{\it {
On the mathematics side, in a correspondingly striking implication it is now known
adinkras, due to the work of \cite{adnkGEO1,adnkGEO2}, may be regarded as 
examples of Grothendieck's ``dessin d' enfant'' and thereby define a class of 
Riemann surfaces.  From this vantage point, the heights of the nodes in an 
adinkra correspond to an integer-valued discrete Morse function defined over 
these Riemann surfaces.}}

In 
1970, the mathematician Banchoff \cite{B1}\footnote{The publications in 
\cite{adnkGEO1,adnkGEO2} pointed toward the relevance of this work.} introduced 
a form of discrete Morse functions for oriented triangular meshes.   For the case 
of the Riemann surfaces associated with adinkras, it is the value of the heights 
of the nodes (made into a piecewise-linear function over the meshes constructed 
by linear extension across the edges and faces of the plaquettes) that are used 
to define the Morse function.  So the height assigned to each node may also
be called the``Banchoff index'' of the node.  
  
Within 4D theories generally not all of the bosonic fields contained within a 
supermultiplet possess the same engineering dimensions.  In a similar manner, 
in the general case, not all fermionic fields contained in a supermultiplet 
possess the same engineering dimensions.  At the level of adinkra graphs, 
the engineering dimension of a particular projection of a space-time field 
determines the literal height at which the node representing that projection 
appears in the graphs.  Fields with the lowest engineering dimension appear 
at the lowest level of an adinkra graph.   Fields of higher engineering dimension 
appear at higher levels in the graph.

A special class of adinkras are those in which all the bosonic nodes in the 
adinkra appear at the same height and all the fermionic nodes in the adinkra 
have the same height (but which is different from the bosonic node height).  
These are called ``valise'' adinkras. In terms of the field theories of which 
such adinkras are their projections, all bosonic fields in the supermultiplet 
possess the same engineering dimension.   A similar statement can be 
made about the fermionic fields of the supermultiplet.

Even before the introduction of adinkras, the phenomenon of ``node lifting''
and "node lowering" had been noticed \cite{ulTRA}.  The effect of lifting nodes in 
adinkra graphs was investigated in \cite{Auto1,Auto2,EH1,EH2,Bowtie,Iso,Adnk2d}.  
Using a Python program, we have computed eigenvalues of ``B-matrices'' 
of all sixteen adinkras corresponding to lifting bosonic nodes of each of the 
36,864 ${\cal GR}(4,4)$ valise adinkra associated with the Coexeter Group 
$BC{}_4$. We refer to the eigenvalues of these B-matrices as Height Yielding 
Matrix Numbers (HYMNs). The main results of this paper are given in 
section~\ref{s:Results} where we classify the B-matrices and their associated 
node-raised adinkras in terms of HYMN equivalence classes. We find that the 
HYMN equivalence classes for valise adinkras are simply the $\chi_{\rm o}
$-equivalence classes originally investigated in~\cite{Gates:2009me} and 
more formally defined in the recent work~\cite{G-1}. We find a bosonic 
node-lifting mirror symmetry in the HYMN equivalence classes for adinkras 
with two bosonic nodes raised.  Source code in the Python language is 
provided in appendix~\ref{a:Python}. A summary of the calculations can 
be found in a \emph{Mathematica} notebook available at the HEPTHools 
\href{https://hepthools.github.io/Data/}{Data Repository} on GitHub.

\section{Review of 36,864 Four-Color Valise Adinkras}\label{s:36864}
  
First, we will briefly review L-matrices. Here we use the 4D, $\cal N$ = 1 chiral 
multiplet as an example. We have in our conventions for 4D quantities,     
\begin{equation}
\label{equ:chiral_mul}
\begin{split}
& {\rm D}{}_a A \,=\, \psi_a ~~,~~ {\rm D}{}_a B \,=\, i(\gamma^5)_a{}^b\psi_b ~~,~~
{\rm D}{}_a F \,=\, (\gamma^{\mu})_a{}^b\pa_{\mu}\psi_b ~~,~~ {\rm D}{}_a G 
\,=\, i(\gamma^5 \gamma^{\mu})_a{}^b\pa_{\mu}\psi_b ~~, \\
& {\rm D}{}_a\psi_b \,=\, i(\gamma^{\mu})_{ab}\pa_{\mu}A-(\gamma^5 
\gamma^{\mu})_{ab}\pa_{\mu}B-iC_{ab}F + (\gamma^5)_{ab}G  ~~.\end{split}
\end{equation}
    
These equations are still valid if we restrict the functions only to be dependent 
on the $t$-coordinate. Under this restriction, we get the 4D, $\cal N$ = 1 
chiral multiplet on the 0-Brane. Define  
\begin{equation}\label{e:CMNodes}
\begin{split}
& \psi_1 \,=\, i\Psi_1 ~~,~~ \ \psi_2 \,=\, i\Psi_2 ~~,~~ \ \psi_3 \,=\, i\Psi_3 ~\,~,~~ \ 
\psi_4 \,=\, i\Psi_4~\,~,\\
& \Phi_1 \,=\, A ~~~~,~~\ \Phi_2 \,=\, B ~~~~,~~ \ \pa_0\Phi_3 \,=\, F ~~,~~ \ \pa_0
\Phi_4 \,=\, G
~~,
\end{split}\end{equation}        
and rewrite (\ref{equ:chiral_mul}) on the 0-Brane in the form
\begin{subequations}\label{e:DI}
\begin{align}
\label{e:DIPhi}
& {\rm D}{}_I \Phi_i = i(L_I)_{i\hat{k}}\Psi_{\hat{k}} ~~~\,~,\\
\label{e:DIPsi}
& {\rm D}{}_I \Psi_{\hat{k}} = (R_I)_{\hat{k}i}\frac{d}{dt}\Phi_i ~~,
\end{align}
\end{subequations}
 where $I = 1, 2, 3, 4$ is the color index. 
        
Note that in (\ref{e:CMNodes}) the projections of the 4D fields $A$ and $B$ are directly related to adinkra 
nodal functions according to $ \Phi_1 = A,\ \Phi_2 = B$, while the projections of the 4D fields 
$F$ and $G$ are related to adinkra nodal functions according to $\pa_0 \Phi_3 = F,\ 
\pa_0\Phi_4 = G$.  That is the adinkra nodal functions $ \Phi_3$ and $ \Phi_4$ are the 
integrals of the projections of the 4D fields $F$ and $G$.  Whenever an integral is used to 
define an adinkra node starting from 
the projection of a 4D field, we refer to this process as ``nodal lowering.''  This
concept was first presented in the work of \cite{ulTRA}.
        
Consequently, we derive the explicit form of L-matrices: 
\begin{equation}
\label{equ:chiral_Lmatrix}
\begin{split}
(L_1)_{i\hat{k}} \,  &= \,  \left(
\begin{matrix}
1 & 0 & 0 & 0 \\
0 & 0 & 0 & -1 \\
0 & 1 & 0 & 0 \\
0 & 0 & -1 & 0 \\
\end{matrix}
\right)~~,\ 
(L_2)_{i\hat{k}} \,=\, \left(
\begin{matrix}
0 & 1 & 0 & 0 \\
0 & 0 & 1 & 0 \\
-1 & 0 & 0 & 0 \\
0 & 0 & 0 & -1 \\
\end{matrix}
\right)~~,\\
(L_3)_{i\hat{k}} \, &= \,  \left(
\begin{matrix}
0 & 0 & 1 & 0 \\
0 & -1 & 0 & 0 \\
0 & 0 & 0 & -1 \\
1 & 0 & 0 & 0 \\
\end{matrix}
\right)~~,\ 
(L_4)_{i\hat{k}} \,=\, \left(
\begin{matrix}
0 & 0 & 0 & 1 \\
1 & 0 & 0 & 0 \\
0 & 0 & 1 & 0 \\
0 & 1 & 0 & 0 \\
\end{matrix}
\right) ~~.
\end{split}
\end{equation}
        
The R-matrices can be obtained by the relation $(R_I) = [(L_I)]^T$, where $T$ superscript 
means transposition. The general algebra for d $\times$ d matrices describing $N$ 
supersymmetries, known as the ${\cal {GR}}({\rm d},N)$ algebra (garden algebra) is
\begin{align}\label{e:GRdN}
\begin{split}
(L_I)_{i\hat{k}} (R_J)_{\hat{k}j} + (L_J)_{i\hat{k}} (R_I)_{\hat{k}j} \,=&\, \,2 \delta_{IJ} \delta_{ij}  ~~,\\
(R_I)_{\hat{i}k} (L_J)_{k\hat{j}} + (R_J)_{\hat{i}k} (L_I)_{k\hat{j}} \,=& \, \,2 \delta_{IJ} \delta_{\hat{i}\hat{j}} ~~.
\end{split}
\end{align}
The $(L_I)_{i\hat{k}}$ and $(R_I)_{\hat{k}i}$  matrices described by Eqs.~(\ref{equ:chiral_Lmatrix}) 
satisfy the ${\cal {GR}}(4,4)$ algebra.

The adinkra diagram for the off-shell chiral multiplet where the associated L-matrices 
appear in Eq.~(\ref{equ:chiral_Lmatrix}) is shown as Figure~\ref{fig:chiral_adinkra}, 
where ${\rm D}{}_1$-green, ${\rm D}{}_2$-violet, ${\rm D}{}_3$-orange, ${\rm D}{}_4$-red.
        
\begin{figure}[htp!]
\includegraphics[width=0.5\textwidth]{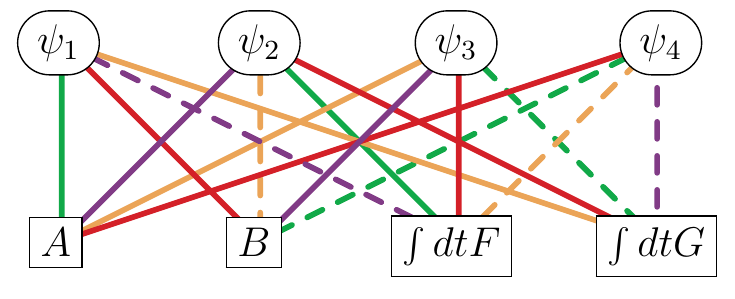}
\centering
\caption{Valise Adinkra for Chiral Supermultiplet with $F$ \& $G$ Nodes Lowered}
\label{fig:chiral_adinkra}
\end{figure}
  	
Next, we will briefly introduce an algorithm to obtain all four-color valise adinkras. The 
recent work~\cite{G-1} studied how to generate all 36,864 four-color valise adinkras from 
only two basis valise adinkras given the titles of ``quaternion adinkras.'' 
    
By definition, $BC_4$ is the group of signed permutations of four elements. $BC_4$ 
can be expressed as plus or minus one times a sign flip element times a permutation 
element times an element of the Vierergruppe. 
\begin{equation}
BC_4^{\pm a\mu A} = \pm H^a S_3^{\mu} \mathcal{V}^A,
\end{equation}
where $H^a$ is a sign flip element and $a = 1, 2, ..., 8$
\begin{equation}
H^a = \{(), (\overline{12}), (\overline{13}), (\overline{23}), (\overline{1}), 
(\overline{2}), (\overline{3}), (\overline{123})\},
\end{equation}
$S_3^{\mu}$ is a permutation element and $\mu = 1, 2, ..., 6$
\begin{equation}
S_3^{\mu} = \{(), (12), (13), (23), (123), (132)\},
\end{equation}
$\mathcal{V}^A$ is an element of the Vierergruppe and $A = 1, 2, 3, 4$
\begin{equation}
\mathcal{V}^A = \{(), (12)(34), (13)(24), (14)(23)\}.
\end{equation}
The explicit matrix forms for these elements are given in \cite{G-1}. As in~\cite{G-1}, we refer 
to sign flips such as $H^a$ as simply \emph{flips} and permutations such as $S_3^{\mu}\mathcal{V
}^A$ as \emph{flops}.

All four-color valise adinkras can be generated by combinations of transformations.
\begin{equation}
 \begin{split}
 (L_I^{\pm a\mu Ab\nu})_i{}^{\hat{j}} \, &= \, (BC_4^{\pm a\mu A})_i{}^j(H^bS_3^{\nu
 })_I{}^J(L_J^{(Q)})_j{}^{\hat{j}},\\
 (\widetilde{L}_I^{\pm a\mu Ab\nu})_i{}^{\hat{j}} \, &= \, (BC_4^{\pm a\mu A})_i{}^j(H^b
 S_3^{\nu})_I{}^J(\widetilde{L}_J^{(\widetilde{Q})})_j{}^{\hat{j}}.
 \end{split}
 \end{equation}
 where $(L_J^{(Q)})_j{}^{\hat{j}}$ and $(\widetilde{L}_J^{(\widetilde{Q})})_j{}^{\hat{j}}$
 denote the L-matrices associated with the quaternionic basis adinkras to be given shortly.
 
The counting is summarized as the following table. The total product is 36,864. Thus, 
there are totally 36,864 four-color valise adinkras.      
\begin{table}[htbp]
\centering
\label{Tab:counting}
\begin{tabular}{|c|c|c|c|c|c|c|c|}
\hline
index & $\pm$ & a & $\mu$ & A & b & $\nu$ & $\sim$ \\
\hline
count & 2     & 8 & 6     & 4 & 8 & 6     & 2  \\
\hline
\end{tabular}
\end{table}

We define $BC_4$ color transformations as $(BC_4^{\pm a\mu A})_I{}^J$, where $I, J$ are 
the color indices; $BC_4$ boson transformations as $(BC_4^{\pm a\mu A})_i{}^j$, where $i, 
j$ are the bosonic indices; $BC_4$ fermion transformations as \newline$(BC_4^{\pm a\mu A
})_{\hat{i}}{}^{\hat{j}}$, where $\hat{i}, \hat{j}$ are the fermionic indices. The definitions of the 
L-matrices associated with the quaternion basis adinkras $Q$ and $\widetilde{Q}$ take the 
explicit
forms,
\begin{equation}
\label{equ:quaternion}
\begin{split}
L_1^{(Q)} \, \, &= \, \, \left(
\begin{matrix}
1 & 0 & 0 & 0 \\
0 & 1 & 0 & 0 \\
0 & 0 & 1 & 0 \\
0 & 0 & 0 & 1 \\
\end{matrix}
\right)~~~~~~,~~\ 
L_2^{(Q)} = \left(
\begin{matrix}
0 & -1 & 0 & 0 \\
1 & 0 & 0 & 0 \\
0 & 0 & 0 & -1 \\
0 & 0 & 1 & 0 \\
\end{matrix}
\right)~~,~~\\
L_3^{(Q)} &= \left(
\begin{matrix}
0 & 0 & -1 & 0 \\
0 & 0 & 0 & 1 \\
1 & 0 & 0 & 0 \\
0 & -1 & 0 & 0 \\
\end{matrix}
\right)~~,~~\ 
L_4^{(Q)} = \left(
\begin{matrix}
0 & 0 & 0 & -1 \\
0 & 0 & -1 & 0 \\
0 & 1 & 0 & 0 \\
1 & 0 & 0 & 0 \\
 \end{matrix}
\right)~~,~~\\
 L_1^{(\widetilde{Q})} \, \, &= \, \, \left(        
\begin{matrix}
1 & 0 & 0 & 0 \\
0 & 1 & 0 & 0 \\
0 & 0 & 1 & 0 \\
0 & 0 & 0 & 1 \\
\end{matrix}
\right)~~~~~,~~\ 
L_2^{(\widetilde{Q})} = \left(
\begin{matrix}
0 & 1 & 0 & 0 \\
-1 & 0 & 0 & 0 \\
0 & 0 & 0 & -1 \\
0 & 0 & 1 & 0 \\
\end{matrix}
\right)~~~, \\
L_3^{(\widetilde{Q})} &= \left(
\begin{matrix}
0 & 0 & -1 & 0 \\
0 & 0 & 0 & -1 \\
1 & 0 & 0 & 0 \\
0 & 1 & 0 & 0 \\
\end{matrix}
\right)~~,~~\ 
L_4^{(\widetilde{Q})} = \left(
\begin{matrix}
0 & 0 & 0 & -1 \\
0 & 0 & 1 & 0 \\
0 & -1 & 0 & 0 \\
1 & 0 & 0 & 0 \\
 \end{matrix}
\right)~~.
\end{split}
\end{equation}

Since the early 1980's \cite{KT} and in later work (see for example \cite{H1,H2,H3,H4})
it has been understood that division algebras play a significant role in the representations 
of spacetime SUSY.  The results in~(\ref{equ:quaternion}) situate these previous observations in the
context of ${\cal {GR}}$(d, $N$) algebras.  In fact the embedding of the
quaternions realized by the matrices in~(\ref{equ:quaternion}) also matches the initial presentation on
 $\cal {GR}$(d, $N$) algebras in \cite{GRana1,GRana2}.  The recursion formulae
 given in these earliest works on Garden Algebras also solve a puzzle.
 
 The division algebras only exists up to the octonions.  This is equivalent to 4D, $\cal N$
 = 2 spacetime SUSY.  Since realizations of SUSY exist beyond 4D, $\cal N$ = 2, the
 puzzle was to ask, ``What replaces the division algebras for the higher theories?"
In the works of \cite{GRana1,GRana2} such extensions were created for all values of $\cal N$ using a recursion formula that realizes Bott periodicity as reviewed in appendix~\ref{a:Recursion}.
		
\section{Adinkra Height Related 4D, \texorpdfstring{$\cal N$}{N} = 1 Minimal Supermultiplets}		

The chiral supermultiplet is only one member of a group of ten such supermultiplets
that lead to adinkras of the same general form as that shown in  Fig.~\ref{fig:chiral_adinkra}.

There are three versions of the 4D, $\cal N$ = 1 chiral supermultiplet related to the 
standard version discussed above by Hodge duality transformation applied to the 
auxiliary fields.  These take the forms shown in equations (\ref{QTd1}), (\ref{QTd2}), 
and (\ref{QTd3}), respectively.  \newline
\noindent
$ {Hodge-Dual~ \#1~Chiral~Supermultiplet: ~(A, \, B, \,  \psi_a , \, {\rm f}_{\mu
 \, \nu \, \rho}, \, G)}$
\be
 \eqalign{
{~~~~} {\rm D}_a A ~&=~ \psi_a  ~~~~~~~~~~~~~~,~~~~
{\rm D}_a B ~=~ i \, (\g^5){}_a{}^b \, \psi_b  ~~~~~~~~~, \cr
{\rm D}_a \psi_b ~&=~ i\, (\g^\m){}_{a \,b}\,  \pa_\m A 
~-~  (\g^5\g^\m){}_{a \,b} \, \pa_\m B ~-~ i \,  \frac 1{3!} \, C_{a\, b} 
\, (\e{}^{\s}{}^{\m}{}^{\n}{}^{\rho} \, \pa_{\s} {\rm f}_{\m
 \, \n \, \rho})  ~+~  (\g^5){}_{ a \, b} G  ~~, \cr
{\rm D}_a {\rm f}_{\m \, \n \, \rho} ~&=~ -\,  (\g^\s){}_a{}^b \,  
\e{}_{\s}{}_{\m}{}_{\n}{}_{\rho} \, \psi_b   
~~~,~~~ 
{\rm D}_a G ~=~ i \,(\g^5\g^\m){}_a{}^b \, \pa_\m \,  
\psi_b  ~~~,} \label{QTd1}
\ee 
${Hodge-Dual~ \#2~Chiral~Supermultiplet: ~(A, \, B, \,  \psi_a , \, F, \, {\rm 
g}_{\m \, \n \, \rho})}$
\be
 \eqalign{
{~~~~} {\rm D}_a A ~&=~ \psi_a  ~~~~~~~~~~~~~~,~~~~
{\rm D}_a B ~=~ i \, (\g^5){}_a{}^b \, \psi_b  ~~~~~~~~~, \cr
{\rm D}_a \psi_b ~&=~ i\, (\g^\m){}_{a \,b}\,  \pa_\m A 
~-~  (\g^5\g^\m){}_{a \,b} \, \pa_\m B ~-~ i \, C_{a\, b} 
\,F  ~+~  \frac 1{3!} \, (\g^5){}_{ a \, b}
\, (\e{}^{\s}{}^{\m}{}^{\n}{}^{\rho} \, \pa_{\s} {\rm g}_{\m
 \, \n \, \rho})  ~~, \cr
{\rm D}_a F ~&=~  (\g^\m){}_a{}^b \, \pa_\m \, \psi_b   
~~~,~~~ 
{\rm D}_a {\rm g}_{\m \, \n \, \rho} ~=~ -\,  (\g^5 \g^\s){}_a{}^b \,  
\e{}_{\s}{}_{\m}{}_{\n}{}_{\rho} \, \psi_b  
 ~~~,
} \label{QTd2}
\ee 
\noindent
$ {Hodge-Dual~ \#3~Chiral~Supermultiplet: ~(A, \, B, \,  \psi_a , \, 
{\rm f}_{\m \, \n \, \rho}, \, {\rm g}_{\m \, \n \, \rho})}$
\be
 \eqalign{
{~~~~} {\rm D}_a A ~&=~ \psi_a  ~~~~~~~~~~~~~~,~~~~
{\rm D}_a B ~=~ i \, (\g^5){}_a{}^b \, \psi_b  ~~~~~~~~~, \cr
{\rm D}_a \psi_b ~&=~ i\, (\g^\m){}_{a \,b}\,  \pa_\m A 
~-~  (\g^5\g^\m){}_{a \,b} \, \pa_\m B \cr
&~~~~-~ i \,  
\frac 1{3!} \, C_{a\, b} \, (\e{}^{\s}{}^{\m}{}^{\n}{}^{\rho} \, 
\pa_{\s} {\rm f}_{\m \, \n \, \rho})
 ~+~  \frac 1{3!} \, (\g^5){}_{ a \, b}
\, (\e{}^{\s}{}^{\m}{}^{\n}{}^{\rho} \, \pa_{\s} {\rm g}_{\m
 \, \n \, \rho})   ~~, \cr
{\rm D}_a {\rm f}_{\m \, \n \, \rho} ~&=~ -\,  (\g^\s){}_a{}^b \,  
\e{}_{\s}{}_{\m}{}_{\n}{}_{\rho} \, \psi_b    
~~~,~~~ 
{\rm D}_a {\rm g}_{\m \, \n \, \rho} ~=~ -\,  (\g^5 \g^\s){}_a{}^b \,  
\e{}_{\s}{}_{\m}{}_{\n}{}_{\rho} \, \psi_b    ~~~.
} \label{QTd3}
\ee  

The engineering dimensions of the three-form gauge fields that appear
above are the same as those of the $A$ and $B$ fields.  So in an adinkra
without node lowering, these three-form fields appear at the same height 
as the $A$ and $B$ fields.  This has implications for how the component 
fields in these supermultiplets are related to the graph shown in Fig.~\ref{fig:chiral_adinkra}.

For the $ {Hodge-Dual~ \#1~Chiral~Supermultiplet}$ one should perform
the replacement in Fig.~\ref{fig:chiral_adinkra}. of the auxiliary fields according to
\be
\int \, dt \, F ~\to~ {\rm f}_{ 1 2 3}  ~~~,~~~ G ~\to~ G  ~~~, 
\label{R_1}
\ee
where f$ {}_{ 1 2 3}$ is the purely spatial component of the Lorentz
3-form  ${\rm f}_{ \mu\nu\rho}$.

For the $ {Hodge-Dual~ \#2~Chiral~Supermultiplet}$ one should perform
the replacement in Fig.~\ref{fig:chiral_adinkra}. of the auxiliary fields according to
\be
F ~\to~ F   ~~~,~~~ 
\int \, dt \, G ~\to~ {\rm g}_{ 1 2 3}  
~~~, 
\label{R_2}
\ee
where g$ {}_{ 1 2 3}$ is the purely spatial component of the Lorentz
3-form  ${\rm g}_{ \mu\nu\rho}$.

For the $ {Hodge-Dual~ \#3~Chiral~Supermultiplet}$ one should perform
the replacement in Fig.~\ref{fig:chiral_adinkra} of the auxiliary fields according to
\be
\int \, dt \, F ~\to~ {\rm f}_{ 1 2 3}  ~~~,~~~
\int \, dt \, G ~\to~ {\rm g}_{ 1 2 3}  
~~~, 
\label{R_3}
\ee
where f$ {}_{ 1 2 3}$ is the purely spatial component of the Lorentz 3-form  ${\rm f
}_{ \mu\nu\rho}$ and g$ {}_{ 1 2 3}$ is the purely spatial component of the Lorentz
3-form  ${\rm g}_{ \mu\nu\rho}$.

The parity transformations and Hodge dual transformations carried out on the 
standard chiral supermultiplet can be extended to vector supermultiplets as well.  
This leads to the supermultiplets described by the four equations seen in
(\ref{QT2}) - (\ref{QTd2b}) according to: \newline  \noindent
$ {Vector~Supermultiplet:~ (A{}_{\mu} , \, \l_b , \,  {\rm d})}$
\be
\eqalign{
{~~~~} {\rm D}_a \, A{}_{\mu} ~&=~  (\gamma_\mu){}_a {}^b \,  \l_b  ~~~, \cr
{\rm D}_a \l_b ~&=~   - \,i \, \fracm 14 ( [\, \gamma^{\mu}\, , \,  \gamma^{\nu} 
\,]){}_a{}_b \, (\,  \pa_\mu  \, A{}_{\nu}    ~-~  \pa_\nu \, A{}_{\mu}  \, )
~+~  (\gamma^5){}_{a \,b} \,    {\rm d} ~~,  {~~~~~~~}  \cr
{\rm D}_a \, {\rm d} ~&=~  i \, (\gamma^5\gamma^\mu){}_a {}^b \, 
\,  \pa_\mu \l_b  ~~~, \cr
}  \label{QT2}
\ee  \vskip 0.12in \noindent
${Axial-Vector~Supermultiplet:~ (U{}_{\mu} , \, {\Tilde \l}_b , \,  {\Tilde {\rm d}}
)}$
\be
\eqalign{
{~~~~} {\rm D}_a \, U{}_{\mu} ~&=~ i\, (\gamma^5 \gamma_\mu){}_a {}^b 
\,  {\Tilde \l}_b  ~~~, \cr
{\rm D}_a {\Tilde \l}_b ~&=~  \, \fracm 14 ( \gamma^5 [\, \gamma^{\mu}\, , \,  
\gamma^{\nu} \,]){}_a{}_b \, (\,  \pa_\mu  \, U{}_{\nu}    ~-~  \pa_\nu 
\, U{}_{\mu}  \, ) ~+~ i \, C{}_{a \,b} \, {\Tilde {\rm d}} ~~,  {~~~~~~~}  \cr
{\rm D}_a \,  {\Tilde {\rm d}} ~&=~  - \, (\gamma^\mu){}_a {}^b \, 
\,  \pa_\mu {\Tilde \l}_b  ~~~, 
\cr}  \label{QT2a}
\ee
${Hodge-Dual~Vector~Supermultiplet:~ (A{}_{\mu} , \, \l_b , \,  {\rm d}
{}_{\mu \, \nu \, \rho} 
)}$
\be
\eqalign{
{~~~~~~~~~~~~~~~~} {\rm D}_a \, A{}_{\mu} ~&=~  (\gamma_\mu){}_a 
{}^b \,  \l_b  ~~~, \cr
{\rm D}_a \l_b ~&=~   - \,i \, \fracm 14 ( [\, \gamma^{\mu}\, , \, \gamma^{\nu} 
\,]){}_a{}_b \, (\,  \pa_\mu  \, A{}_{\nu}    ~-~  \pa_\nu \, A{}_{\mu}  \, )
~+~  \frac 1{3!} \, (\gamma^5){}_{ a \, b} \, (\epsilon{}^{\s}{}^{\mu}{}^{\nu}
{}^{\rho} \, \pa_{\s} {\rm d}{}_{\mu \, \nu \, \rho} )  ~~,  {~~~~~~~}  \cr
{\rm D}_a \, {\rm d} ~&=~  i \, (\gamma^5\gamma^\mu){}_a {}^b \, 
\,  \pa_\mu \l_b  ~~~, \cr
}  \label{QTd2a}
\ee 
\noindent
${Hodge-Dual~ Axial-Vector~Supermultiplet:~ (A{}_{\mu} , \, {\Tilde \l}
{}_b , \,  {\Tilde {\rm d}}{}_{\mu \, \nu \, \rho} )}$
\be
\eqalign{
{~~~~} {\rm D}_a \, U{}_{\mu} ~&=~ i\, (\gamma^5 \gamma_\mu){}_a {}^b 
\,  {\Tilde \l}_b  ~~~, \cr
{\rm D}_a {\Tilde \l}_b ~&=~  \, \fracm 14 ( \gamma^5 [\, \gamma^{\mu}\, , \,  
\gamma^{\nu} \,]){}_a{}_b \, (\,  \pa_\mu  \, U{}_{\nu}    ~-~  \pa_\nu 
\, U{}_{\mu}  \, ) ~+~ i \,  \frac 1{3!} \, C{}_{a \,b} \, \, (\epsilon{}^{\s}{}^{\mu}
{}^{\nu}{}^{\rho} \, \pa_{\s} {\Tilde {\rm d}}{}_{\mu \, \nu \, \rho} )  ~~,    \cr
{\rm D}_a \,  {\Tilde {\rm d}} ~&=~  - \, (\gamma^\mu){}_a {}^b \, \pa_\mu 
 {\Tilde \l}_b    ~~~, \cr
}  \label{QTd2b}
\ee

One can continue by applying Hodge transformations to the propagating
physical bosons of the chiral supermultiplet.  There are theories associated
with performing the duality on either the scalar or pseudoscalar in the
supermultiplet.  In the former case this leads to the tensor supermultiplet
shown in (\ref{QT3}), while in the latter case the duality transformation
leads to the axial-tensor supermultiplet shown in (\ref{QT3a}).

 \noindent
$ {Tensor~Supermultiplet: ~(\varphi, \, B{}_{\mu \, \nu }, \,  \chi_a )}$
\be
 \eqalign{
{\rm D}_a \varphi ~&=~ \chi_a  ~~~,~~~~~ 
{\rm D}_a B{}_{\mu \, \nu } ~=~ -\, \fracm 14 ( [\, \gamma_{\mu}
\, , \,  \gamma_{\nu} \,]){}_a{}^b \, \chi_b  ~~~, \cr
{\rm D}_a \chi_b ~&=~ i\, (\gamma^\mu){}_{a \,b}\,  \pa_\mu \varphi 
~-~  (\gamma^5\gamma^\mu){}_{a \,b} \, \e{}_{\mu}{}^{\r \, \s \, \t}
\pa_\r B {}_{\s \, \t}~~, {~~~~~~~~~~~~~~\,~~}
}  \label{QT3}
\ee
 \noindent
$ {Axial-Tensor~Supermultiplet: ~( {\Tilde {\varphi}}, \, 
C{}_{\mu \, \nu }, \,  {\Tilde {\chi}} {}_a )}$
\be
 \eqalign{
{\rm D}_a {\Tilde {\varphi}} ~&=~ - i \, (\gamma^\mu){}_{a}{}^{b} \, {\Tilde {\chi}}{}_b  ~~~,~~~~~ 
{\rm D}_a C{}_{\mu \, \nu } ~=~ -i\, \fracm 14  ( \gamma^5 [\, \gamma_{\mu}
\, , \,  \gamma_{\nu} \,]){}_a{}^b \,  {\Tilde {\chi}}{}_b  ~~~, \cr
{\rm D}_a  {\Tilde {\chi}}{}_b ~&=~ -\, (\gamma^5 \gamma^\mu){}_{a \,b}\,  \pa_\mu 
 {\Tilde {\varphi}}
~-~  i \, (\gamma^\mu){}_{a \,b} \, \e{}_{\mu}{}^{\r \, \s \, \t}
\pa_\r C{}_{\s \, \t}~~, {~~~~~~~~~~~~~~\,~~}
}  \label{QT3a}
\ee
From these results, it is seen there are no explicit auxiliary fields in
either case.  Implicitly, such fields are present in the gauge two-forms
contained in each supermultiplet.    

One might be tempted to continue the process of performing a Hodge
duality on the remaining scalar or pseudoscalar in either of the two
version of a tensor supermultiplet above.  This was carried out in
a little known work~\cite{Freedman:1977pa} by Freedman many years ago.  However,
the field theory that results from this ``dualization'' is distinguished from
all the other supermultiplets described in the chapter.  Namely, one
can use the transformation laws of the ten supermultiplets given in
(\ref{equ:chiral_mul}), (\ref{QTd1}), $\dots$, (\ref{QT3a}) to show 
the condition
\be
\{ \, {\rm D}_a ~,~ {\rm D}_b \, \} ~=~ i \, 2 (\g{}^{\mu}){}_{a \, b} \pa{}_{\m}
\ee
is satisfied (up to gauge transformations) on every component field...except
in the case where Hodge dualization is carried on scalar and pseudoscalar
in the two tensor supermultiplets above.

So all toll, there are ten minimal {\em{off-shell}} 4D, $\cal N$ = 1 supermultiplets.
    
\section{HYMNs: Height Yielding Matrix Numbers}

Valise adinkras have the advantage of being the simplest type of adinkra.  However,
we know that when attempting to use adinkra-based arguments to re-construct higher 
dimensional field theory representations, the simplicity of valise adinkras interacts with
the requirement of the realization of Lorentz symmetry in the higher dimensions
\cite{EH1,EH2,Bowtie}.  There can arise obstructions as pointedly noted in \cite{Bowtie,Adnk2d}
where the conditions for selecting which adinkras can be used as a basis for the
re-construction of 2D world sheet supersymmetric theories.  These works demonstrate
that the shape of a non-valise adinkra is an important attribute to knowing what higher
dimensional physics relates to which adinkra systems.

In the works of \cite{Auto1,Auto2,Iso} two apparently distinct approaches were presented
as a way to characterize the shape of non-valise adinkras.  In particular, the second of
these approaches in \cite{Iso} makes explicit the appearance of a particular matrices derived 
from the L-matrices and R-matrices.  These derived matrices can be called the ``left B-matrix'' 
and the ``right B-matrix.''  Due to their relations to the work in \cite{B1} this is a highly 
appropriate name.  The eigenvalues of the B-matrices relate to the shape of all adinkras 
including therefore, non-valise ones.  In the following discussion we are going to define 
the B-matrices in a manner that is slightly different from the work \cite{Iso}. Our new method has the benefit of categorizing raised-node adinkras in a way that is manifestly color-independent, or supersymmetric charge-independent, in contrast to the methods of~\cite{Iso} which are manifestly color-dependent.

In the work of \cite{Iso} a proposition was made for when two adinkras are isomorphic that involved the two matrices constructed similarly to $B_L =  L_{N}R_{N-1}L_{N-2}\cdots$ and $B_R = R_{N}L_{N-1}R_{N-2}\cdots$ that contained color-dependent height parameters $\beta_I$. It was proposed that if two adinkras are isomorphic, their chromocharacters and their eigenvalues for $B_L$ and $B_R$ are the same. The chromocharacter of a non-valise adinkra is defined, up to a normalization, as the trace of its $B_L$ matrix after setting all $\beta_I$ to one. This indicates an importance in developing eigenvalue equivalence classes for non-valise adinkras, which we wish to do in color-independent way without use of the $\beta_I$. We have recently developed a new procedure to define such eigenvalue equivalence classes.  In this section, we will briefly describe this procedure using the 
simplest non-trivial adinkras. We begin by revisiting the valise adinkra for the chiral multiplet, 
where we notice the mass dimensions of the fields are:
\begin{align}
[ \Phi_i ] \,=\, 1~~~,~~~[\Psi_{\hat{i}}] \,=\, 3/2    ~~~.
\end{align}
For any valise adinkra, the bosons will all have the same mass dimension and the fermions 
will all have the same mass dimension that is one-half higher than the bosons. We can raise 
the nodes associated with the auxiliary bosons $F$ and $G$ via the raising operator $M$, 
defined below along with its inverse, the lowering operator $M^{-1}$
\begin{align}
M = \begin{pmatrix}
1 & 0 & 0 & 0 \\
0 & 1 & 0 & 0 \\
0 & 0 & \pa_0 & 0 \\
0 & 0 & 0 & \pa_0
\end{pmatrix}~~~,~~~   M^{-1} = & \begin{pmatrix}
1 & 0 & 0 & 0 \\
0 & 1 & 0 & 0 \\
0 & 0 & \int~dt & 0 \\
0 & 0 & 0 & \int~dt 
\end{pmatrix}    ~~~.
\end{align}
The $M$ operator raises nodes by performing the field redefinition:
\begin{align}
\tilde{\Phi} \,=\, & M \Phi \,=\, \begin{pmatrix}
A \\
B \\
F \\
G \\
\end{pmatrix}    ~~~.
\end{align}
This results in the following mass dimensions for the tilded fields $\tilde{\Phi}_i$:
\begin{align}
[\tilde{\Phi}_1] \,=\, [\tilde{\Phi}_2] \,=\, 1~~~,~~~[\tilde{\Phi}_3] \,=\, [\tilde{\Phi}_4] 
\,=\, 2~~~.
\end{align}
Operating on Eq.~(\ref{e:DIPhi}) from the left with $M$ results in the following:
\begin{align}\label{e:DItPhiproof}
{\rm D}{}_I M\Phi \, &=\, i M L_I \Psi   ~~~, \cr
 {\rm D}{}_I \tilde{\Phi} \,&=\, i \tilde{L}_I \tilde{\Psi} ~~~~~~,
\end{align}
where we have defined $\tilde{\Psi} = \Psi$ and $\tilde{L}_I$ as follows
\begin{align}
\tilde{L}_I  \,=\, M L_I ~~~.
\end{align}
Inserting $M^{-1} M = I$ into Eq. (\ref{e:DIPsi}) results in
\begin{align}\label{e:DItPsiproof}
{\rm D}{}_I \Psi \,&=\, R_I M^{-1} M \Phi  ~~~, \cr
{\rm D}{}_I \tilde{\Psi} \,&=\, \tilde{R}_I \tilde{\Phi}  ~~~~~~~~~~\,~~.
\end{align}
We write the results in Eqs.~(\ref{e:DItPhiproof}) and~(\ref{e:DItPsiproof}) succinctly 
as the transformation laws for $\tilde{\Phi}$ and $\tilde{\Psi}$:   
\begin{subequations}\label{e:DItSummary}
\begin{align}
{\rm D}{}_I \tilde{\Phi} \,&=\, i \tilde{L}_I \tilde{\Psi}   ~~~, \\
{\rm D}{}_I \tilde{\Psi} \,&=\, \tilde{R}_I \tilde{\Phi}  ~~~~.
\end{align}
\end{subequations}
It is straightforward to show that $\tilde{L}_I$ and $\tilde{R}_I$ satisfy the garden 
algebra, Eq.~(\ref{e:GRdN}), written instead in terms of these tilded matrices. The 
adinkra matrices $\tilde{L}_I$ and $\tilde{R}_I$ now have height information, in 
the derivatives and integrals, and can be expressed as the three-level adinkra 
in Fig.~\ref{f:CMExt}
\begin{figure}[htp!]
\includegraphics[width=0.5\textwidth]{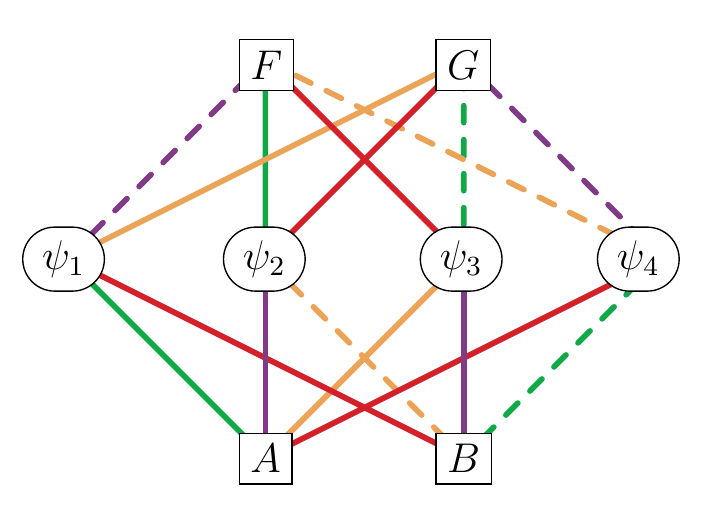}
\centering
\caption{Adinkra for Chiral Supermultiplet without $F$ \& $G$ Nodes Lowered}
\label{f:CMExt}
\end{figure}
   
To investigate such node liftings of adinkras, it will be advantageous to forget the 
integral/derivative nature of lowering/raising nodes and instead simply use mass 
parameters to keep track the lowered/raised nodes. We make the substitution
\begin{align}
\pa_0 F \to m F~~~,~~~\pa_0 G \to m G  ~~~.
\end{align}For instance, for the chiral multiplet $M$ and its inverse could have been 
defined instead as
\begin{align}
M(m) = \begin{pmatrix}
1 & 0 & 0 & 0 \\
0 & 1 & 0 & 0 \\
0 & 0 & m & 0 \\
0 & 0 & 0 & m
\end{pmatrix}   ~~~,
\end{align}
and Eqs.~(\ref{e:DItPhiproof}) through (\ref{e:DItSummary}) would have followed the 
same as before. 
        
Next, we generalize these node raising matrices to arbitrarily sized ${\rm d} \times {\rm d}$ adinkras. The ${\cal GR}({\rm d},N)$ 
SUSY transformation laws are
\begin{align}
\label{e:DIPhiGRdN}
{\rm D}{}_I \Phi \, \,=\, \, & i L_I \Psi  ~~~, \\
\label{e:DIPsiGRdN}
{\rm D}{}_I \Psi \,  \,=\,\, & R_I  \dot{\Phi} ~~~~,
\end{align} 
where a dot above a field indicates a time derivative: $\dot{\Phi} = d\Phi/dt$. The $L_I$ 
and $R_I$ matrices realize the closed $N=4$ SUSY algebra for $d$ bosons and $d$
fermions as the ${\cal GR}({\rm d},N)$ algebra
\begin{align}
\label{e:LRGRdN}
L_I R_J + L_J R_I \,=&\, \, 2 \delta_{IJ} I_{\rd\times\rd}  ~~~, \\
\label{e:RLGRdN}
R_I L_J + R_J L_I \,=&\,\, 2 \delta_{IJ}  I_{\rd\times \rd} ~~~, 
\end{align}
with $I_{\rd\times \rd}$ the $\rd \times \rd$ identity matrix.
    
We define the node raising operator $M(m,w)$ that acts on an arbitrary number of d
bosons as
\begin{align}\label{e:pword}
M(m,w) \equiv & \begin{pmatrix}
m^{p_1} & 0 & 0 & \dots & 0 \\
0 & m^{p_1} & 0 & \dots & 0 \\
0 & 0 & m^{p_3} & \dots & 0 \\
\vdots & \vdots & \vdots & \ddots & 0 \\
0 & 0 & 0 & \dots & m^{p_{\rd}}
\end{pmatrix} ~~~,
\\ {~} \nonumber \\
\label{e:word}
w \equiv & p_1 2^{0} + p_2 2^1 + p_3 2^{2} + \dots + p_{\rd} 2^{\rd-1}~~~,~~~\text{with}~p_i 
= 0,1
\end{align}
where the word parameter $w$ is as in~\cite{permutadnk}. The node raising operator 
$M(m,w)$ has the following combination properties
\begin{align}
\label{e:MProp0}
M(1,w) ~=~ I{}_{\rd \times \rd}   ~~~, \\
\label{e:MProp1}
M(m,w) M(\mu,w) ~=~& M(m\mu,w) ~~~, \\
\label{e:MProp2}
M(m,w_1)M(m,w_2) ~=~& M(m,w_1 + w_2) ~~~, \\
\label{e:MProp3}
[M(m,w)]^{-1} ~=~& M(m^{-1},w) ~~~,\\
\label{e:MProp4}
M(m,w)[M(\mu,w)]^{-1} ~=~& M(m/\mu,w)  ~~~,
\end{align}
We denote a raised node boson collection as $\Phi(m,w)$ 
\begin{align}\label{e:PhiPsiRedef}
\Phi(m,w) ~=~ &  M(m,w) \Phi ~~~.
\end{align}Multiplying Eq.~(\ref{e:DIPhiGRdN}) from the left by $M(m,w)$ and inserting $I{}_{
\rd \times \rd} = M(\mu,w) [M(\mu,w)]^{-1}$ into the right hand side of Eq.~(\ref{e:DIPsiGRdN}) 
results in
\begin{align}\label{e:DIRedef0}
\begin{split}
{\rm D}{}_I M(m,w) \Phi \,=&\, \, i M(m,w) L_I \Psi~~~,~~~ \cr
{\rm D}{}_I \Psi \,=&\,\,  R_I  [M(\mu,w)]^{-1} M(\mu,w) \dot{\Phi} 
= ~R_I M(\mu^{-1},w) M(\mu,w)\dot{\Phi} ~~~,
\end{split}
\end{align}
We now make the matrix redefinitions
\begin{align}\label{e:LRraised}
L_I(m,w) = M(m,w) L_I~~~,~~~R_I(\mu,w) =& R_I  M(\mu^{-1},w)
\end{align}
which along with the field redefinitions in Eq.~(\ref{e:PhiPsiRedef}) reduces Eqs.~(\ref{e:DIRedef0}) 
to
\begin{align}\label{e:DIRedef1}
{\rm D}{}_I \Phi(m,w)  = & i L_I(m,w) \Psi~~~,~~~ {\rm D}{}_I \Psi =  R_I(\mu,w)\dot{\Phi}(\mu,w)
\end{align}
It is important to make clear the meaning of the matrices $L_I(m,w)$ and $ R_I(\mu,w)$ as these
are the "deformed" version of the $L_I$ and $ R_I$ for adinkras where the some number of
bosonic and fermionic nodes have been lifted from the corresponding level in a valise adinkra
within the context of $BC{}_4$ related adinkras.  It will be necessary to modify our formalism
in the future to handle the cases where nodes can be lifted to a height that is greater than
one about that which the node appears in a corresponding valise adinkra.

The redefined matrices $L_I(m,w)$ and $R_I(\mu,w)$ satisfy the $GR(d,N)$ algebra in the 
$\mu \to m$ limit:
\begin{align}
L_I(m,w) R_J(\mu,w) +  L_J(m,w) R_I(\mu,w) =& \,M(m,w)(L_IR_J + L_JR_I) M(\mu^{-1},w) \cr
=&\, 2 \delta_{IJ} M(m,w) M(\mu^{-1},w) \cr
=& \, 2 \delta_{IJ} M(m/\mu,w) \cr
 \to &\,  2 \delta_{IJ} I{}_{\rd \times \rd} ~~~,~~~ \text{for}~\mu \to m
\end{align}
where going from the second to third line we have made use of the property in Eq.~(\ref{e:MProp1}) and in going from the third to last line have made use of the property in Eq.~(\ref{e:MProp4}). The same results holds in the $\mu \to m$ limit for the $R_IL_J + R_JL_I$ 
algebra in Eq.~(\ref{e:RLGRdN}), though the details are different:
\begin{align}
R_I(\mu,w) L_J(m,w) + R_J(\mu,w) L_I(m,w) =&\, R_I M(\mu^{-1},w)M(m,w) L_J + R_J 
M(\mu^{-1},w)M(m,w) L_I \cr
=& \, R_I M(m/\mu,w) L_J + R_J M(m/\mu,w) L_I \cr
\to& \, R_I L_J + R_J L_I = 2 \delta_{IJ} I{}_{\rd \times \rd}~~~,\text{for}~\mu \to m~~~.
\end{align}

In order to describe adinkras uniquely, the color dependent block matrix $C_I$ associated 
with the $I$-th color is defined as
\begin{equation}\label{e:C}
C_I = 
\begin{pmatrix}
0 & L_I \\
R_I & 0 \\
\end{pmatrix} ~~~.
\end{equation}
Since every path can be covered by looking at the N distinct color paths from a boson or 
fermion, we can define the $B$ matrix by multiplying all of the color matrices
\begin{equation}
B = C_NC_{N-1}C_{N-2}\cdots C_1 ~~~.
\end{equation}
If $N$ is odd, 
\begin{equation}
B = \left(
\begin{matrix}
0 & L_N R_{N-1} \cdots L_1 \\
R_N L_{N-1} \cdots R_1& 0 \\
\end{matrix}
\right) \, \equiv \, 
\left(
\begin{matrix}
0 & B_L \\
B_R & 0 \\
\end{matrix}
\right) ~~~,
\end{equation}
and if $N$ is even,
\begin{equation}\label{e:Banchoff}
B = \left(
\begin{matrix}
L_N R_{N-1} \cdots R_1 & 0 \\
0 & R_N L_{N-1} \cdots L_1 \\
\end{matrix}
\right)\, \equiv  \, 
\left(
\begin{matrix}
B_L & 0 \\
0 & B_R \\
\end{matrix}
\right) ~~~.
\end{equation}
    
Any reader familiar with our work in \cite{Iso} will recognize these definitions from that 
work.    We here make the additional identification that the non-vanishing upper left 
hand entry in the last expression of (\ref{e:Banchoff}) is what we mean by the left B- matrix 
and accordingly the non-vanishing lower right hand entry in the last expression 
of (\ref{e:Banchoff}) is the right B-matrix. We define Height Yielding Matrix Numbers 
(HYMNs) as the eigenvalues of these matrices. It also is of note that in the case of 
odd values of $N$, the square of the matrix in (4.30) can be used to define the left  and 
right B-matrices.
    
Permuting fermionic or bosonic nodes of a valise adinkra does not influence its HYMNs. We prove this in the following. We can assign that if we relabel fermionic nodes, 
the permutation matrix is $P_F$; if we relabel bosonic nodes, the permutation matrix 
is $P_B$. Thus, after relabeling, 
\begin{equation}
L_{I}\rightarrow P_BL_{I}P_F,\ R_{I}\rightarrow P^{T}_FR_{I}P^{T}_B ~~~,
\end{equation}
Consequently, when $N$ is odd,
\begin{equation}
\begin{split}
&L_{N}R_{N-1}\cdots L_{1} \, \rightarrow \, P_BL_{N}P_F P^{T}_FR_{N-1}P^{T}_B...P_BL_{1}P_F 
\, =\,  P_B(L_{N}R_{N-1}\cdots L_{1})P_F ~~~,\\
&R_{N}L_{N-1}\cdots R_{1} \, \rightarrow \, P^{T}_FR_{N}P^{T}_B P_BL_{N-1}P_F\cdots P^{
T}_FR_{1}P^{T}_B \, = \, P^{T}_F(R_{N}L_{N-1}\cdots R_{1})P^{T}_B ~~~.
\end{split}
\end{equation}
When $N$ is even,
\begin{equation}
\begin{split}
&L_{N}R_{N-1}\cdots R_{1} \, \rightarrow \, P_BL_{N}P_F P^{T}_FR_{N-1}P^{T}_B\cdots P^{T
}_FR_{1}P^{T}_B \, = \, P_B(L_{N}R_{N-1}\cdots R_{1})P^{T}_B ~~~,\\
&R_{N}L_{N-1}\cdots L_{1} \, \rightarrow \, P^{T}_FR_{N}P^{T}_B P_BL_{N-1}P_F\cdots P_BL_{
1}P_F \, = \, P^{T}_F(R_{N}L_{N-1}\cdots L_{1})P_F ~~~~\,.
\end{split}
\end{equation}
For all $N$, we can obtain
\begin{equation}
B\rightarrow 
\left(
\begin{matrix}
P_B & 0 \\
0 & P^{T}_F \\
\end{matrix}
\right)B
\left(
\begin{matrix}
P^{T}_B & 0 \\
0 & P_F \\
\end{matrix}
\right) ~~~.
\end{equation}
Thus, eigenvalues of $B$ will not be changed after relabeling nodes, although eigenvalues 
of $L_{N}...L_{1}$ and $R_{N}...R_{1}$ may be changed when $N$ is odd. 

We conjecture that the HYMNs of an adinkra carry all information about its shape isomorphisms. Specifically, we conjecture that two adinkras are isomorphic if and only if their HYMNs are the same.

The parameter $\chi{}_{\rm o}$ defines the two equivalence classes for $\mathcal{GR}(4,4)$ valise 
adinkras with respect to signed permutations of bosonic and fermionic nodes~\cite{Gates:2009me}.
The general definition of $\chi{}_{\rm o}$ for adinkras with arbitrary number of colors $N$ is 
\cite{Auto1,Auto2}
\begin{equation}
\chi{}_{\rm o} ~=~ \frac{1}{\rd_{\rm min}(N)}\frac{1}{N!}\e^{I_1I_2\cdots I_N}\, Tr(L_{I_1}R_{I_2}
\cdots L/R_{I_N})
~~~,
\end{equation}
where the last matrix in the trace will be $L_{I_N}$ if $N$ is odd and $R_{I_N}$ if $N$ is even. 
The quantity $\rd_{\rm min}(N)$ is a function of $N$ first identified in the original works on
Garden Algebras \cite{GRana1,GRana2}
\begin{equation}\label{e:dmin}
\rd_{\rm min}(N) ~=~ \left\{
\begin{aligned}
&2^{\frac{N-1}{2}},\ N\equiv 1, 7\ \mod(8)\\
&2^{\frac{N}{2}},\ N\equiv 2, 4, 6\ \mod(8)\\
&2^{\frac{N+1}{2}},\ N\equiv 3, 5\ \mod(8)\\
&2^{\frac{N-2}{2}},\ N\equiv 8\ {~~~} \mod(8)\\
\end{aligned} 
\right.    ~~~~.
\end{equation}

So far, the effect of lifting nodes on adinkra isomorphisms is generally unknown. In order to study 
this effect, we are developing a symbolic program of which we give examples in the 
following subsections.

\subsection{Raised Boson Adinkras for \texorpdfstring{${\cal {GR}}(2,2)$}{{\cal {GR}}(2,2)}}
Figure~\ref{f:GR22L0} shows one of the simplest two-color adinkras. The adinkra matrices 
for this adinkra are the following where $L_1$-green and $L_2$-purple:
\begin{equation}\label{e:L0}
\begin{split}
L_1 &= \left(
\begin{matrix}
1 & 0 \\
0 & 1 \\
\end{matrix}
\right) ~~,\ 
R_1 = \left(
\begin{matrix}
1 & 0 \\
0 & 1 \\
\end{matrix}
\right) ~~~~,\\
L_2 &= \left(
\begin{matrix}
0 & -1 \\
1 & 0 \\
\end{matrix}
\right),\ 
R_2 = \left(
\begin{matrix}
0 & 1 \\
-1 & 0 \\
\end{matrix}
\right) ~~.
\end{split}
\end{equation}
      
\begin{figure}[htp!]
\centering
\includegraphics[width=0.2\textwidth]{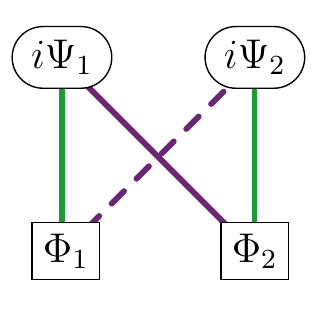}
\caption{An example of a $\mathcal{GR}(2,2)$ valise adinkra.}
\label{f:GR22L0}
\end{figure}
    
In Fig.~\ref{f:GR22L0} for instance, we can raise the first bosonic node with 
$M(m,1)$, the second with $M(m,2)$ and both with $M(m,3) = M(m,1)M(m,2)$. 
These matrices and the corresponding raised boson collections $\Phi(m,w)$ are   
\begin{subequations}
\begin{align}\label{e:GR22RaisedNodes}
\Phi(m,1) \, =& \,
\begin{pmatrix}
m \Phi_1 \\
\Phi_2 
\end{pmatrix}
~~~,~~~ M(m,1) \,=\,  \begin{pmatrix}
m & 0 \\
0 & 1
\end{pmatrix} ~~~, \\
\Phi(m,2) \,=& \,  \begin{pmatrix}
\Phi_1 \\
m \Phi_2 \\
\end{pmatrix}
~~~,~~~ M(m,2) = \begin{pmatrix}
1 & 0 \\
0 & m
\end{pmatrix}  ~~~~, \\
\Phi(m,3) =&  \begin{pmatrix}
m \Phi_1 \\
m \Phi_2 \\
\end{pmatrix} 
~~~,~~~ M(m,3) = \begin{pmatrix}
m & 0 \\
0 & m
\end{pmatrix} ~~.
\end{align}
\end{subequations}
The resulting adinkras for each of these three cases are given in Fig.~\ref{f:GR22Raised}. 

\begin{figure}[!htbp]
\centering
\subfigure[Valise]{
\label{f:GR22Valise}
\includegraphics[width=0.2\textwidth]{GR22p111Labelled}}
\quad 
\subfigure[First boson raised.]{
\label{f:GR22FirstRaised}
\includegraphics[width=0.2\textwidth]{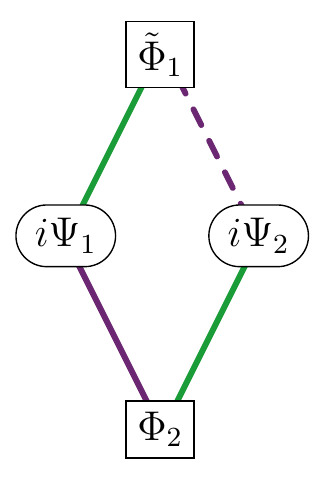}}
\quad 
\subfigure[Second boson raised.]{
\label{f:GR22SecondRaised}
\includegraphics[width=0.2\textwidth]{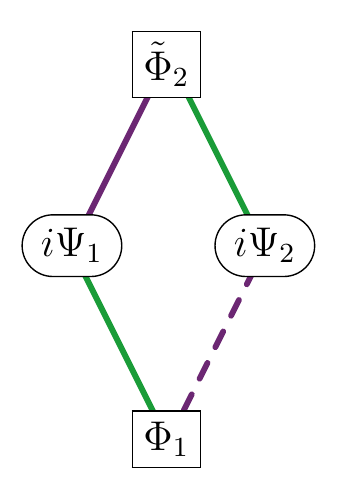}}
\quad 
\subfigure[Both bosons raised.]{
\label{f:GR22BothRaised}
\includegraphics[width=0.2\textwidth]{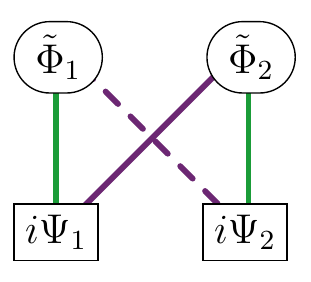}}
\caption{Valise and raised versions of the ${\cal {GR}}(2,2)$ adinkra of Fig.~\ref{f:GR22L0}. 
The tilded  bosons are identified as  $\tilde{\Phi}_i = m \Phi_i$. }
\label{f:GR22Raised}
\end{figure}
According to Eq.~(\ref{e:LRraised}), the adinkra matrices associated with Fig.~\ref{f:GR22Raised} 
and nodal redefinitions in Eq.~(\ref{e:GR22RaisedNodes}) are as follows. The adinkra matrices 
corresponding to the leftmost adinkra in Fig.~\ref{f:GR22Raised} with boson one raised are
\begin{equation}\label{e:LRGR22Raised1}
\begin{split}
L_1(m,1) \, &=  \, \left(
\begin{matrix}
m & 0 \\
0 & 1 \\
\end{matrix}
\right) ~~,\ 
R_1(\mu,1) = \left(
\begin{matrix}
\mu^{-1} & 0 \\
0 & 1 \\
\end{matrix}
\right)~~\,~~,\\
L_2(m,1) \, &=  \, \left(
\begin{matrix}
0 & -m \\
1 & 0 \\
\end{matrix}
\right),\ 
R_2(\mu,1) = \left(
\begin{matrix}
0 & 1 \\
 -\mu^{-1} & 0 \\
\end{matrix}
\right) ~~.
\end{split}
\end{equation}
The adinkra matrices corresponding to the middle adinkra in Fig.~\ref{f:GR22Raised} with 
boson two raised are
\begin{equation}\label{e:LRGR22Raised2}
\begin{split}
L_1(m,2) \, &=  \, \left(
\begin{matrix}
1 & 0 \\
0 & m \\
\end{matrix}
\right)~~~,\ 
R_1(\mu,2) = \left(
\begin{matrix}
1 & 0 \\
0 & \mu^{-1} \\
\end{matrix}
\right) ~~~~~~,\\
L_2(m,2) \, &=  \, \left(
\begin{matrix}
0 & -1 \\
m & 0 \\
\end{matrix}
\right) ~,\ 
R_2(\mu,2) = \left(
\begin{matrix}
0 & \mu^{-1} \\
-1 & 0 \\
\end{matrix}
\right) ~~\,~.
\end{split}
\end{equation}
The adinkra matrices corresponding to the rightmost adinkra in Fig.~\ref{f:GR22Raised} with 
both bosons raised are
\begin{equation}\label{e:LRGR22Raised12}
\begin{split}
L_1(m,3) \, &=  \, \left(
\begin{matrix}
m & 0 \\
0 & m \\
\end{matrix}
\right) ~~~\,~,\ 
R_1(\mu,3) = \left(
\begin{matrix}
\mu^{-1} & 0 \\
0 & \mu^{-1} \\
\end{matrix}
\right) ~~~~,\\
L_2(m,3) \, &=  \, \left(
\begin{matrix}
0 & -m \\
m & 0 \\
\end{matrix}
\right)  ~~,\ 
R_2(\mu,3) = \left(
\begin{matrix}
0 & \mu^{-1} \\
-\mu^{-1} & 0 \\
\end{matrix}
\right) ~~.
\end{split}
\end{equation}
The two color $C_I$ matrix and $B$ matrix given by Eqs.~(\ref{e:C}) and~(\ref{e:Banchoff}) are
\begin{equation}\label{e:GR22Bdef}
C_I = 
\begin{pmatrix}
0 & L_I \\
R_I & 0 \\
\end{pmatrix}~~~,~~~    B = C_2C_1 = \begin{pmatrix}
L_2R_1 & 0 \\
0 & R_2 L_1
\end{pmatrix}~~~.
\end{equation}
The eigenvalues of $B$ are invariant with respect to fermionic and bosonic nodal transformations 
as described previously for the arbitrary $N$ color case. With respect to color transformations, the 
eigenvalues of $B$ are invariant with respect to even numbers of flips and flops, but odd numbers 
of flips and flops negate all eigenvalues. This is because of the ${\cal {GR}}(2,2)$ algebra relation 
$L_2 R_1 = - L_1 R_2$. The consequences of raising nodes are unknown as to the eigenvalues, 
so we next turn to the tabulation of all ${\cal {GR}}(2,2)$ adinkras, valise and raised.
    
The four images in Figure 4 illustrate a number of points
in the abilities of adinkras to encode supermultiplet in higher
dimensions.  If these figures are interpreted to describe
supermultiplets that solely depend on a single temporal
coordinate, then we see there are three different types of
supermultiplets. 
 
The valise supermultiplet in Fig.~\ref{f:GR22Valise} corresponds to the 1D
projection of a 2D, $\cal N$ = (2,0) SUSY heterotic scalar
supermultiplet \cite{Saka,UniD} that resides on the world
sheet of a string. 
 
The two adinkras in Figs.~\ref{f:GR22FirstRaised} and~\ref{f:GR22SecondRaised} correspond to the 1D projection
of a scalar supermultiplet on the world sheet of a string with
(1,1) SUSY.  In fact, these seemingly distinct supermultiplets are
related one to the other.  One need only simultaneously exchange
the purple and green colors of the links and then change the
sign of the fermion $\Psi{}_2$ to see this.  The supermultiplet
corresponds to the string coordinates on the world sheet. On
the other hand, if one starts with the adinkra in Fig~\ref{f:GR22Valise},
performs a simultaneous exchange of the green and purple
colors of the links, followed by changing the sign of the
 fermion $\Psi{}_2$, it does {\em {not}} become the adinkra
in Fig.~\ref{f:GR22BothRaised} .

Finally, the adinkra in Fig.~\ref{f:GR22BothRaised} corresponds to the projection
of a 2D, $\cal N$ = (2,0) SUSY heterotic fermion supermultiplet
that resides on the world sheet of a string \cite{Saka,UniD}. 
 
 So straight away the power of adinkras to unify distinct SUSY
representations is made apparent.
 
We should also mention the two theories with only 2D, $\cal
N$ = (2,0) SUSY cannot be made into theories that live on a
2D world sheet that possesses $\cal N$ = (1,1) SUSY.  Many
years ago, a result was derived that may be described as the
``no two color ambidextrous bow-tie'' theorem\cite{Bowtie}. 
There it was shown that if one attempted to interpret the
supermultiplets in Figs.~\ref{f:GR22Valise} and~\ref{f:GR22BothRaised} in the context of $\cal
N$ = (1,1) SUSY on the world sheet.  Lorentz invariance must
be broken.  This same result was more rigorously and mathematically
extended in the works of \cite{THB,Adnk2d}.
    
In total, there are 16 distinct ${\cal {GR}}(2,2)$ valise adinkras as shown in Fig.~\ref{f:AllGR22}. 
These are tabulated in terms of sign flips $H_1^a$ and permutations $S_2^{\mu}$ where $a = 
1,2$ and $\mu = 1,2$
\begin{align}
H_1^a = \{ (), (\overline{1}) \}~~~,~~~S_2^\mu = \{ (), (12) \} ~~~.
\end{align}
Analogous to the $BC_4$ notation of~\cite{G-1} that was summarized in Sec.~\ref{s:36864}, all 
16 adinkras in Fig.~\ref{f:AllGR22} are succinctly labeled as
\begin{align}\label{e:LAllGR22}
L^{\pm a\mu b}_I =&  \pm H_1^a S_2^\mu L_I^{(0)} H_1^b~~~,~~~R^{\pm a\mu b}_I = (L^{\pm 
a\mu b}_I)^T  ~~~,
\end{align}
where $L_I^{(0)}$ corresponds to the adinkra in Fig.~\ref{f:GR22L0} with adinkra matrices 
as in Eq.~(\ref{e:L0}) and the $T$ superscript means transpose.  This is analogous to the 
$BC_4$ notation of~\cite{G-1} that was summarized in Sec.~\ref{s:36864}.
\begin{figure}
\centering
\subfigure[$L_I^{+111} = L_I^{(0)}$]{
\label{f:Lp111}
\includegraphics[width = 0.2\textwidth]{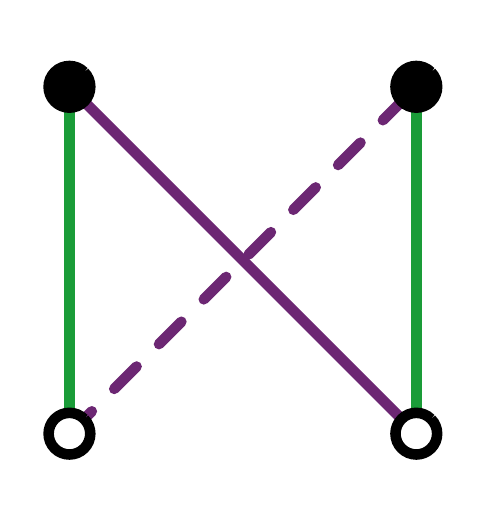}}
\qquad
\subfigure[$L_I^{+211}$]{
\label{f:Lp211}
\includegraphics[width = 0.2\textwidth]{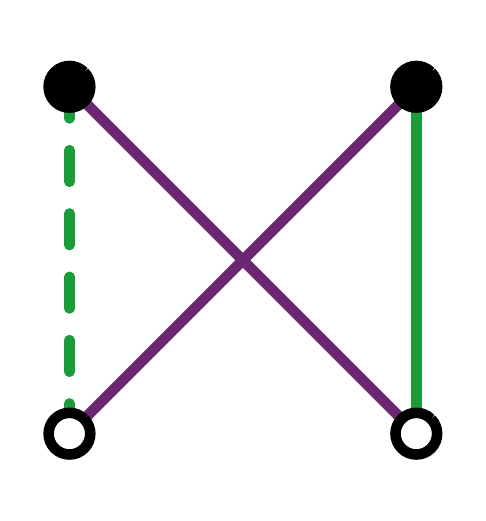}}
\qquad
\subfigure[$L_I^{-211}$]{
\label{f:Lm211}
\includegraphics[width = 0.2\textwidth]{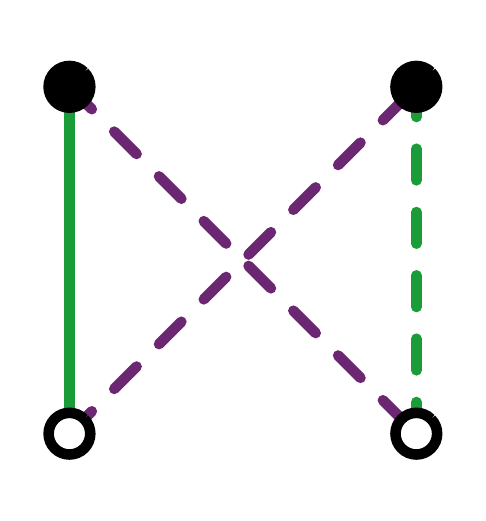}}
\qquad
\subfigure[$L_I^{-111}$]{
\label{f:Lm111}
\includegraphics[width = 0.2\textwidth]{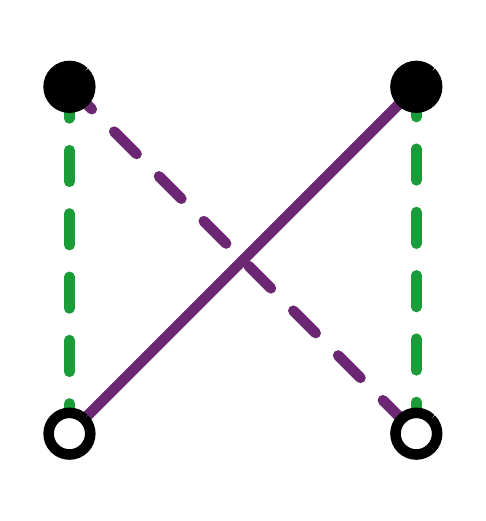}}
\subfigure[$L_I^{+121}$]{
\label{f:Lp121}
\includegraphics[width = 0.2\textwidth]{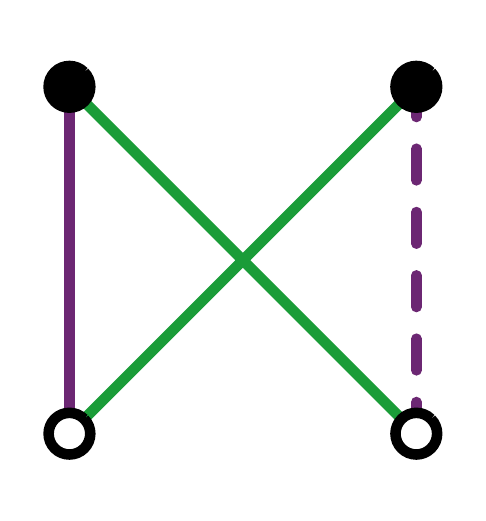}}
\qquad
\subfigure[$L_I^{+221}$]{
\label{f:Lp221}
\includegraphics[width = 0.2\textwidth]{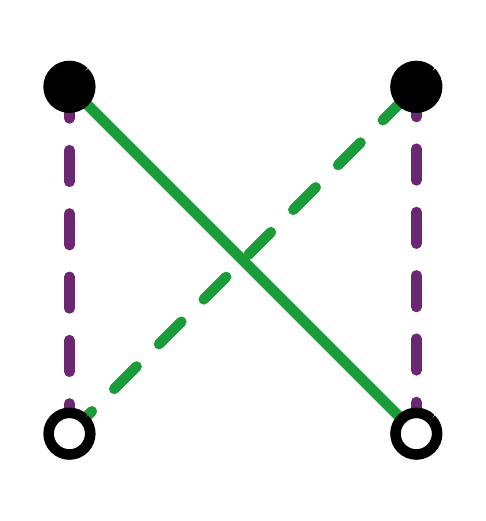}}
\qquad
\subfigure[$L_I^{-221}$]{
\label{f:Lm221}
\includegraphics[width = 0.2\textwidth]{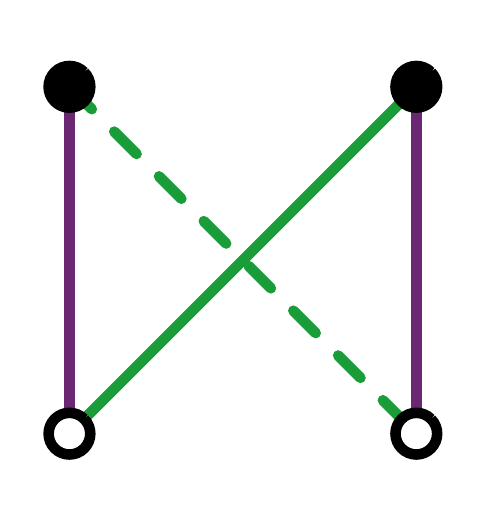}}
\qquad
\subfigure[$L_I^{-121}$]{
\label{f:Lm121}
\includegraphics[width = 0.2\textwidth]{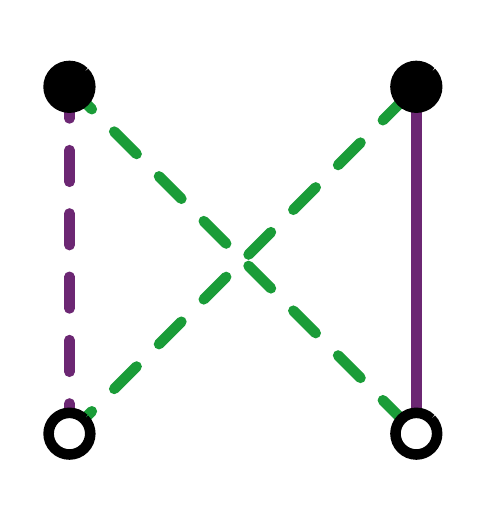}}
\subfigure[$L_I^{+112}$]{
\label{f:Lp112}
\includegraphics[width = 0.2\textwidth]{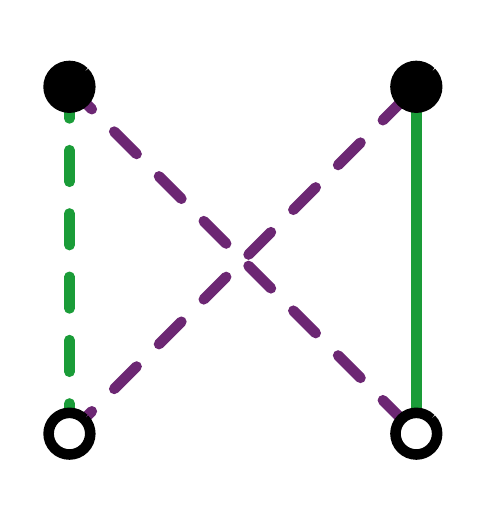}}
\qquad
\subfigure[$L_I^{+212}$]{
\label{f:Lp212}
\includegraphics[width = 0.2\textwidth]{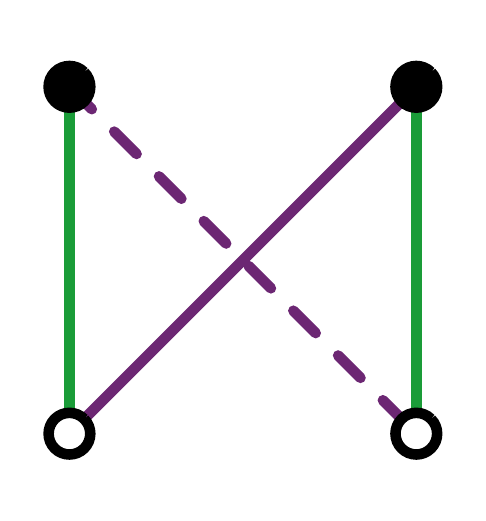}}
\qquad
\subfigure[$L_I^{-212}$]{
\label{f:Lm212}
\includegraphics[width = 0.2\textwidth]{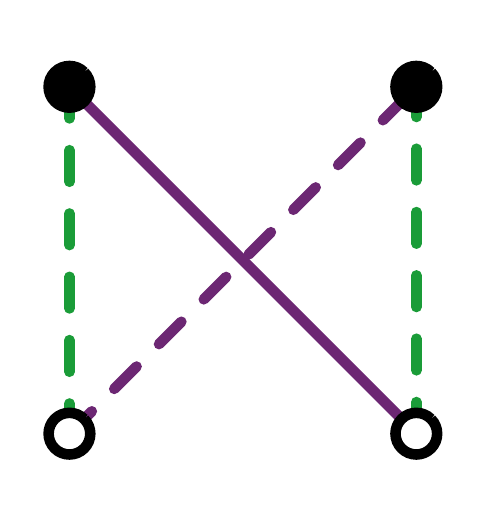}}
\qquad
\subfigure[$L_I^{-112}$]{
\label{f:Lm112}
\includegraphics[width = 0.2\textwidth]{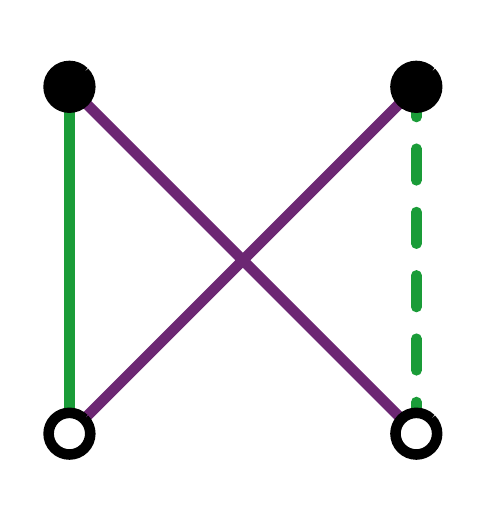}}
\subfigure[$L_I^{+122}$]{
\label{f:Lp122}
\includegraphics[width = 0.2\textwidth]{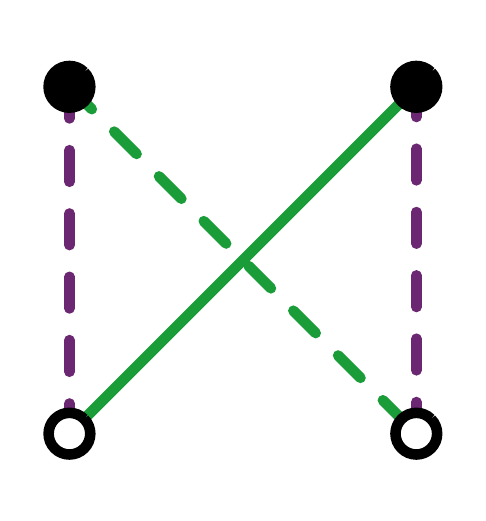}}
\qquad
\subfigure[$L_I^{+222}$]{
\label{f:Lp222}
\includegraphics[width = 0.2\textwidth]{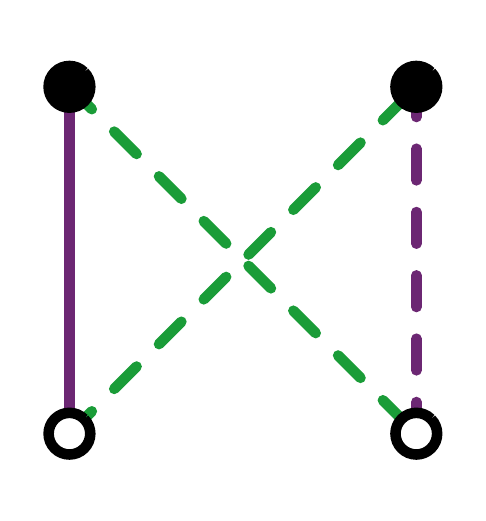}}
\qquad
\subfigure[$L_I^{-222}$]{
\label{f:Lm222}
\includegraphics[width = 0.2\textwidth]{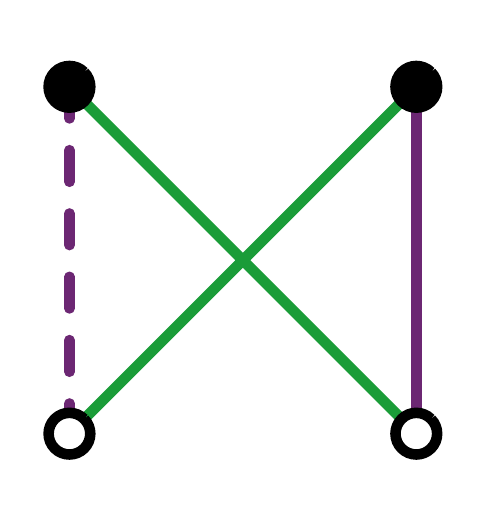}}
\qquad
\subfigure[$L_I^{-122}$]{
\label{f:Lm122}
\includegraphics[width = 0.2\textwidth]{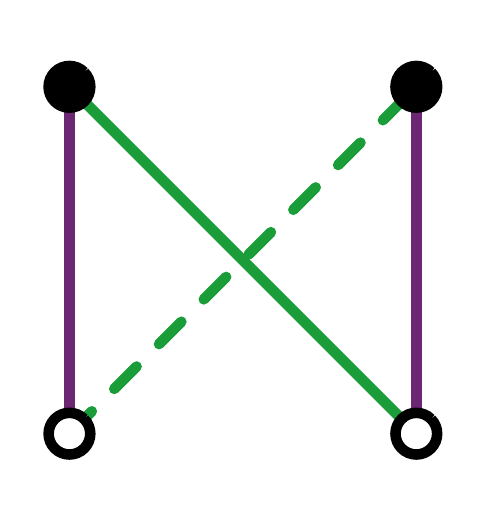}}
\caption{All 16 ${\cal {GR}}(2,2)$ valise adinkras. The white nodes are bosons, 
the black fermions, in numerical left to right order as in Fig.~\ref{f:GR22L0}.}
\label{f:AllGR22}
\end{figure}
Similar to the $BC_4$ case, there are isometries of ${\cal {GR}}(2,2)$ adinkras 
that reduce the total number of distinct adinkras from the 64 total possible $BC_2$ 
boson $\times BC_2$ fermion transformations to the 16 distinct adinkras in 
Fig.~\ref{f:AllGR22}. These isometries are:
\begin{subequations}
\begin{align} 
L_I =& \mathcal{X} L_I \mathcal{Y} \\
(\mathcal{X},\mathcal{Y}) \in & \left\{ 
\begin{array}{l}
((),()) \\
((\Bar{1})(12),(\Bar{1})(12)) \\
((\Bar{2})(12),(\Bar{2})(12)) \\
(\Bar{12},(\Bar{12})) 
\end{array}
\right.
\end{align}
\end{subequations}
These isometries are the reason that Eq.~(\ref{e:LAllGR22}) can be written as 
it is, in terms of only flips on boson one, flops of bosons one and two, flips of 
fermion one, and an overall factor of plus or minus one. Altogether, these are 
the 16 possibilities that take a given starting ${\cal {GR}}(2,2)$ valise adinkra 
to all other possible ${\cal {GR}}(2,2)$ valise adinkras. There is nothing special 
about our choice for $L_I^{(0)} = L_I^{+111}$, it was chosen at random.
    
Next, we construct the 64 possible raisings of each of the 16 ${\cal {GR}}(2,2)$ 
adinkras as in Eq.~(\ref{e:LRraised}):
\begin{align}
\label{equ:64lifting}
L^{\pm a\mu b}_I(m,w)= M(m,w)L^{\pm a\mu b}_I~~~,~~~R^{\pm a\mu b}_I(
\mu,w) = R^{\pm a\mu b}_I M(\mu^{-1},w)  ~~~.
\end{align}
As in Eq.~(\ref{e:GR22Bdef}), we construct $B^{\pm a\mu b}$
\begin{align}
B^{\pm a\mu b}(m,\mu,w) =& \begin{pmatrix}
L^{\pm a\mu b}_2(m,w) R^{\pm a\mu b}_1(\mu,w)& 0 \\
0 & R^{\pm a\mu b}_2(\mu,w)L^{\pm a\mu b}_1(m,w)
\end{pmatrix}~~~\text{no $a,\mu, b$ sum}.
\end{align}
    
This inspires us to calculate the eigenvalues of the following six matrices (no $a, 
\mu,  b$ sum):
\begin{align}\label{e:LR1}
& L^{\pm a\mu b}_2(m,w) R^{\pm a\mu b}_1(\mu,w) = -  L^{\pm a\mu b}_1(m,w) 
R^{\pm a\mu b}_2(\mu,w)   ~\,~~, \\
\label{e:LR2}
&     M(m/\mu,w) =  L^{\pm a\mu b}_1(m,w) R^{\pm a\mu b}_1(\mu,w) = L^{\pm a
\mu b}_2(m,w) R^{\pm a\mu b}_2(\mu,w) ~~~, \\
\label{e:RL1}
&  R^{\pm a\mu b}_2(\mu,w) L^{\pm a\mu b}_1(m,w) ~~~,~~~ R^{\pm a\mu b}_1(\mu,w) 
L^{\pm a\mu b}_2(m,w) ~~~, \\
&  R^{\pm a\mu b}_1(\mu,w) L^{\pm a\mu b}_1(m,w) ~~~,~~~R^{\pm a\mu b}_2(\mu,w) 
L^{\pm a\mu b}_2(m,w)  ~~~.
\end{align} 
The calculation of these eigenvalues amounts to solving the following characteristic 
polynomials (no $a, \mu,  b$ sum):
\begin{align}\label{e:GR22CharP1}
0=& \, \det\left(L^{\pm a\mu b}_2(m,w) R^{\pm a\mu b}_1(\mu,w) \pm 
\lambda I_{2  \times 2}  \right)~~~,  \\
\label{e:GR22CharP2}
0=&\, \det\left(M(m/\mu,w) - \lambda I_{2  \times 2}  \right) ~~~, \\
\label{e:GR22CharP3}
\, 0=&  \,\det\left(R^{\pm a\mu b}_2(\mu,w) L^{\pm a\mu b}_1(m,w) - \lambda I_{2  
\times 2}  \right) ~~~, \\
\label{e:GR22CharP4}
\, 0=&  \,\det\left(R^{\pm a\mu b}_1(\mu,w) L^{\pm a\mu b}_2(m,w) - \lambda I_{2 
 \times 2}  \right)  ~~~, \\
\label{e:GR22CharP5}
\, 0=&  \,\det\left(R^{\pm a\mu b}_1(\mu,w) L^{\pm a\mu b}_1(m,w) - \lambda I_{2  
\times 2}  \right) ~~~,   \\
\label{e:GR22CharP6}
\, 0=&  \,\det\left(R^{\pm a\mu b}_2(\mu,w) L^{\pm a\mu b}_2(m,w) - \lambda I_{2  
\times 2}  \right)  ~~~.
\end{align} Owing to Eq.~(\ref{e:LR1}), the plus sign in Eq.~(\ref{e:GR22CharP1}) 
corresponds to the eigenvalues of $L^{\pm a\mu b}_1(m,w) R^{\pm a\mu b}_2(\mu,w)$.

\subsection{Raised Boson Adinkras for \texorpdfstring{${\cal {GR}}(4,4)$}{{\cal {GR}}(4,4)}}
  
Consider lifting one, two, three, and four bosonic nodes in a four-color valise adinkra with 
four bosons and four fermions. From Sec.~\ref{s:36864}, we have known there are 36,864 
\texorpdfstring{${\cal {GR}}(4,4)$}{{\cal {GR}}(4,4)} valise adinkras in total. Similar to the analysis 
in \texorpdfstring{${\cal {GR}}(2,2)$}{{\cal {GR}}(2,2)} case, we define node raising operators: 
\begin{equation}
M(m,1) = \left(
\begin{matrix}
m & 0 & 0 & 0 \\
0 & 1 & 0 & 0 \\
0 & 0 & 1 & 0 \\
0 & 0 & 0 & 1\\
\end{matrix}
\right)~~,\ 
M(\mu^{-1},1) = \left(
\begin{matrix}
\mu^{-1} & 0 & 0 & 0 \\
0 & 1 & 0 & 0 \\
0 & 0 & 1 & 0 \\
0 & 0 & 0 & 1\\
\end{matrix}
\right)~~,\ 
\end{equation} 

\begin{equation}
M(m,2) = \left(
\begin{matrix}
1 & 0 & 0 & 0 \\
0 & m & 0 & 0 \\
0 & 0 & 1 & 0 \\
0 & 0 & 0 & 1\\
\end{matrix}
\right) ~~,\ 
M(\mu^{-1},2) = \left(
\begin{matrix}
1 & 0 & 0 & 0 \\
0 & \mu^{-1} & 0 & 0 \\
0 & 0 & 1 & 0 \\
0 & 0 & 0 & 1\\
\end{matrix}
\right) ~~,\ 
\end{equation} 

\begin{equation}
M(m,4) = \left(
\begin{matrix}
1 & 0 & 0 & 0 \\
0 & 1 & 0 & 0 \\
0 & 0 & m & 0 \\
0 & 0 & 0 & 1\\
\end{matrix}
\right) ~~,\ 
M(\mu^{-1},4) = \left(
\begin{matrix}
1 & 0 & 0 & 0 \\
0 & 1 & 0 & 0 \\
0 & 0 & \mu^{-1} & 0 \\
0 & 0 & 0 & 1\\
\end{matrix}
\right)  ~~,\ 
\end{equation} 

\begin{equation}
M(m,8) = \left(
\begin{matrix}
1 & 0 & 0 & 0 \\
 0 & 1 & 0 & 0 \\
 0 & 0 & 1 & 0 \\
 0 & 0 & 0 & m\\
\end{matrix}
\right) ~~,\ 
M(\mu^{-1},8) = \left(
\begin{matrix}
1 & 0 & 0 & 0 \\
0 & 1 & 0 & 0 \\
0 & 0 & 1 & 0 \\
0 & 0 & 0 & \mu^{-1} \\
\end{matrix}
\right) ~~,\ 
\end{equation} 
    
For example, if we lift the i-th bosonic node, then the corresponding $L_I$ and $R_I$ 
for the lifted adinkra are
\begin{equation}
L_{I}(m,w) = M(m,w) L_{I} ~,~ R_{I}(\mu,w) = R_{I}M(\mu^{-1},w) ~~~,
\end{equation} 
where $w = 2^{i-1}$.  Lifting more than one bosonic nodes can be described in the 
similar way. For example, the $L_{I}$and $R_{I}$ matrices after lifting the $i-th$ 
and $j-th$ bosonic nodes are:
\begin{equation}
\begin{split}
L_I(m,w) & ~=~ M(m,2^{i-1})M(m,2^{j-1})L_{I} ~~~\,~, \\
R_I(\mu,w) & ~=~ R_{I}M(\mu^{-1},2^{i-1})M(\mu^{-1},2^{j-1})~~.
\end{split}
\end{equation}
where node raising operators commute to each other, which means that lifting the 
$i-th$ node first then the $j-th$ node describes the same thing as lifting the $j-th$ 
node first then the $i-th$ node. In order to study isomorphic properties of non-valise 
adinkras, we study the $B$ matrix
\begin{equation}
\label{equ:B_matrix}
\begin{split}
B \, &=  \, \left(
\begin{matrix}
0 & L_{4} \\
R_{4} & 0 \\
\end{matrix}
\right)
\left(
\begin{matrix}
0 & L_{3} \\
R_{3} & 0 \\
\end{matrix}
\right)
\left(
\begin{matrix}
0 & L_{2} \\
R_{2} & 0 \\
\end{matrix}
\right)
\left(
\begin{matrix}
0 & L_{1} \\
R_{1} & 0 \\
\end{matrix}
\right) ~~~, \\
& = \left(
\begin{matrix}
L_{4}R_{3}L_{2}R_{1} & 0 \\
0 & R_{4}L_{3}R_{2}L_{1} \\
\end{matrix}
\right) ~~~.
\end{split}
\end{equation}
        

        
Next, we construct the 589,824 possible raisings of each of the 36,864 \texorpdfstring{$
{\cal {GR}}(4,4)$}{{\cal {GR}}(4,4)} valise adinkras. Then we have
\begin{align}
B_L(m,\mu,w) ~ &= ~ L_{4}(m,w)R_{3}(\mu,w)L_{2}(m,w)R_{1}(\mu,w)  ~~~, \\
B_R(m,\mu,w) ~&=~ R_{4}(\mu,w)L_{3}(m,w)R_{2}(\mu,w)L_{1}(m,w) ~~~, 
\end{align}
where the word parameter $w$ can be $0, 1, 2, 3, \dots 15$. 
The code (Listing~\ref{script}) in the appendix~\ref{a:Python} is the program to calculate the HYMNs, i.e. the eigenvalues
of $B_L(m,\mu,w)$ and $B_R(m,\mu,w)$, for all 589,824 possible raisings. A summary using 
a \emph{Mathematica} notebook can be found at the HEPTHools \href{https://hepthools.github.io/Data/}{Data Repository} on GitHub.
           
 \section{Results}\label{s:Results}
  
Let us define the mass ratio parameter $\rho$ to connect $m$ and $\mu$,
\begin{equation}
\rho ~\equiv~ \frac{m}{\mu}~~.
\end{equation}
We shall find that the eigenvalues of the matrices described previously will depend only on this 
ratio parameter.

\subsection{HYMNs and Other Eigenvalues for \texorpdfstring{${\cal {GR}}(2,2)$}{{\cal {GR}}(2,2)}} As shown 
by Eq.~(\ref{equ:64lifting}), there are 64 possible ${\cal GR}(2,2)$ valise and raised adinkras. 
The eigenvalues of $L_1(m,w)R_1(\mu,w)$, $L_2(m,w)R_1(\mu,w)$, $R_1(\mu,w) L_1(m,w)$, 
$R_1(\mu,w)L_2(m,w)$, $R_2(\mu,w)L_1(m,w)$, and \\
$R_2(\mu,w)L_2(m,w)$ define eigenvalue equivalence classes of 
\texorpdfstring{${{\cal {GR}}}(2,2)$}{{\cal {GR}}(2,2)} adinkras.  We find
\begin{itemize}
\item all sixteen valise \texorpdfstring{${\cal {GR}}(2,2)$}{{\cal {GR}}(2,2)} 
adinkras give the same eigenvalues: \\
$L_1(m,w)R_1(\mu,w):\, \{ 1~,~1 \}$;~~$L_2(m,w)R_1(\mu,w):\, \{ i~,~-i\}$;
~~$R_1(\mu,w)L_1(m,w):\, \{ 1~,~1 \}$;\\
$R_1(\mu,w)L_2(m,w):\, \{ i~,~-i\}$;~~$R_2(\mu,w)L_1(m,w):\, \{i~,~-i \}$;~~ 
$R_2(\mu,w)L_2(m,w):\, \{ 1~,~1 \}$
\item all \texorpdfstring{${\cal {GR}}(2,2)$}{{\cal {GR}}(2,2)} adinkras with one 
boson raised give the same eigenvalues:\\
$L_1(m,w)R_1(\mu,w):\, \{ 1~,~\rho \}$;~~  $L_2(m,w)R_1(\mu,w):\, \{ i\sqrt{\rho
}~,~-i\sqrt{\rho}\}$;~~$R_1(\mu,w)L_1(m,w):\, \{ 1~,~\rho \}$;\\
$R_1(\mu,w)L_2(m,w):\, \{ i\sqrt{\rho}~,~-i\sqrt{\rho}\}$;~~$R_2(\mu,w)L_1(m,w):
\, \{i\sqrt{\rho}~,~-i\sqrt{\rho} \}$; ~~$R_2(\mu,w)L_2(m,w):\, \{ 1~,~\rho\}$
\item  all \texorpdfstring{${\cal {GR}}(2,2)$}{{\cal {GR}}(2,2)} adinkras with two 
bosons raised give the same eigenvalues: \\
$L_1(m,w)R_1(\mu,w):\, \{ \rho~,~\rho \}$;~~  $L_2(m,w)R_1(\mu,w):\, \{ i\rho
~,~-i\rho\}$;~~$R_1(\mu,w)L_1(m,w):\, \{ \rho~,~\rho \}$;\\
$R_1(\mu,w)L_2(m,w):\, \{ i\rho~,~-i\rho\}$;~~$R_2(\mu,w)L_1(m,w):\, \{i\rho
~,~-i\rho \}$;~~ $R_2(\mu,w)L_2(m,w):\, \{ \rho~,~\rho \}$
\end{itemize}

These eigenvalue equivalence classes are summarized in Table~\ref{Tab:GR22-summary}. 
    
\begin{table}[htp!]
\centering
\caption{Eigenvalue Equivalence Classes of \texorpdfstring{${\cal {GR}}
(2,2)$}{{\cal {GR}}(2,2)}\label{Tab:GR22-summary}}
\begin{tabular}{|c|c|c|c|c|}
\hline
Type & w                                                                       & Total Number & Eigenvalue
Equivalence Class & \begin{tabular}[c]{@{}l@{}}\# of Adinkras \\ in Each Class\end{tabular} \\ \hline
Valise Adinkras & 0                                                               & 16           & all             & 16                   \\ \hline
\begin{tabular}[c]{@{}l@{}}One Boson \\ Raised Adinkras\end{tabular} & 
1,2 & 32           & all           & 32                   \\ \hline
\begin{tabular}[c]{@{}l@{}}Two Bosons\\ Raised Adinkras\end{tabular} &
3 & 16           & all           & 16                   \\ \hline
\end{tabular}
\end{table}
    
\subsection{HYMNs for \texorpdfstring{${\cal {GR}}(4,4)$}{{\cal {GR}}(4,4)}}
    
For ${\cal GR}(4,4)$, we focus specifically on HYMNs, i.e. the eigenvalues for $B_L(\rho,w)$ 
and $B_R(\rho,w)$,  as no additional information was found for ${\cal GR}(2,2)$ in for instance the eigenvalues of $L_2(m,w) R_1(m,w)$ that was not already present in the HYMNs. As $B_L(m,\mu,w)$ and $B_R(m,\mu,w)$ are found to depend only on $\rho$, we refer to these 
here as $B_L(\rho,w)$ and $B_R(\rho,w)$.  We have calculated the HYMNs for all 589,824 possible valise and raised adinkras.  We summarize the HYMN equivalence classes below.
      
\begin{enumerate}
\item 36,864 \texorpdfstring{${\cal {GR}}(4,4)$}{{\cal {GR}}(4,4)} valise adinkras split 
into two classes: $\chi_{\rm o} = \pm 1$, in each class adinkras give the same HYMNs\\
$B_L(\rho,w):\, \chi_{\rm o}\,\{ 1,1,1,1 \} $;~~~ $B_R(\rho,w):\, \chi_{\rm o}\,\{ -1,-1,-1,-1 \}$
 \item 147,456 \texorpdfstring{${\cal {GR}}(4,4)$}{{\cal {GR}}(4,4)} adinkras with one 
 boson raised split into two classes: $\chi_{\rm o} = \pm 1$, in each class adinkras give 
 the same HYMNs\\
 $B_L(\rho,w):\, \chi_{\rm o}\,\{ 1,1,\rho,\rho \} $;~~~ $B_R(\rho,w):\, \chi_{\rm o}\,\{ -1,-1,-\rho,-\rho \}$
\item 227,184 \texorpdfstring{${\cal {GR}}(4,4)$}{{\cal {GR}}(4,4)} adinkras with two bosons 
raised split into six eigenvalue equivalence classes $EB_1^{(\chi_{\rm o})}$, $EB_2^{(\chi_{\rm o})}$, 
and $EB_3^{(\chi_{\rm o})}$.  In each class, the matrices $B_L(\rho,w)$ and $B_R(\rho,w)$ have 
the same HYMNs \\ 
$EB_1^{(\chi_{\rm o})}$: $B_L(\rho,w): \chi_{\rm o}\,\{ \rho,\rho,\rho,\rho \} $; $B_R(\rho,w):\, \chi_{\rm o}\,
\{ -1,-1,-\rho^2,-\rho^2 \}$ \\
$EB_2^{(\chi_{\rm o})}$: $B_L(\rho,w):\, \chi_{\rm o}\,\{ 1,1,\rho^2,\rho^2 \} $; $B_R(\rho,w):\, \chi_{\rm o}\,\{ 
-\rho,-\rho,-\rho,-\rho \}$ \\
$EB_3^{(\chi_{\rm o})}$:  $B_L(\rho,w):\, \chi_{\rm o}\,\{ \rho,\rho,\rho,\rho \} $; $B_R(\rho,w):\, \chi_{\rm o}\,\{ 
-\rho,-\rho,-\rho,-\rho \}$ 
\item 147,456 \texorpdfstring{${\cal {GR}}(4,4)$}{{\cal {GR}}(4,4)} adinkras  with three 
bosons raised split into two classes: $\chi_{\rm o} = \pm 1$, in each class adinkras give 
the same HYMNs\\
$B_L(\rho,w):\, \chi_{\rm o}\,\{ \rho,\rho,\rho^2,\rho^2 \} $; $B_R(\rho,w):\, \chi_{\rm o}\,\{ -\rho,-\rho,
-\rho^2,-\rho^2 \}$
\item 36,864 \texorpdfstring{${\cal {GR}}(4,4)$}{{\cal {GR}}(4,4)} adinkras with all four bosons 
raised split into two classes: $\chi_{\rm o} = \pm 1$, in each class adinkras give the same 
HYMNs\\
$B_L(\rho,w):\, \chi_{\rm o}\,\{ \rho^2,\rho^2,\rho^2,\rho^2 \} $; $B_R(\rho,w):\, \chi_{\rm o}\,\{ -\rho^2,
-\rho^2,-\rho^2,-\rho^2 \}$
\end{enumerate}

These eigenvalue equivalence classes are summarized in Table~\ref{Tab:GR44-summary}.  

\begin{table}[!htbp]
\centering
\caption{HYMN Equivalence Classes of \texorpdfstring{${\cal {GR}}(
4,4)$}{{\cal {GR}}(4,4)}\label{Tab:GR44-summary}}
\begin{tabular}{|c|c|c|c|c|}
\hline
\multirow{2}{*}{Type} & \multirow{2}{*}{w} & \multirow{2}{*}{Total Number} &
HYMN Equivalence Classes & \begin{tabular}[c]{@{}l@{}}\# of Adinkras \\ in Each Class
\end{tabular} \\ 
\hline
Valise Adinkras  & 0 & 36,864 & $\chi_{\rm o} = \pm 1$ & 18,432 \\ \hline
\begin{tabular}[c]{@{}l@{}}One Boson \\ Raised Adinkras\end{tabular} & 1,2,4,8            
& 147,456  & $\chi_{\rm o} = \pm 1$   & 73,728  \\ \hline
\begin{tabular}[c]{@{}l@{}}Two Bosons\\ Raised Adinkras\end{tabular} & 3,5,6,9,10,12      
& 221,184 & $EB_1^{(\pm)}$, $EB_2^{(\pm)}$ , $EB_3^{(\pm)}$ & 36,864  \\ \hline
\begin{tabular}[c]{@{}l@{}}Three Bosons\\ Raised Adinkras\end{tabular} & 7,11,13,14         
& 147,456  & $\chi_{\rm o} = \pm 1$ & 73,728  \\ \hline
\begin{tabular}[c]{@{}l@{}}Four Bosons\\ Raised Adinkras\end{tabular}  & 15                 
& 36,864 & $\chi_{\rm o} = \pm 1$  & 18,432 \\ \hline
\end{tabular}
\end{table}
}     

There is an interesting relationship between adinkras within the $EB_i^{(\pm)}$ equivalence 
classes for two raised bosons. For a given valise adinkra, raising bosons one and two are in 
the same equivalence class as raising bosons three and four. The same holds for raising one 
and three or two and four as well as for one and four or two and three. In terms of words, for a 
given valise adinkra raising two bosons with either the word code $w$ or $15-w$ will lead to 
the same equivalence class. Specifically, these word pairs are $(3,12)$, $(5,10)$ and $(6,9)$. 
It is important to note that this does not mean that all adinkras raised with word $w=3$ have 
the same HYMNs as all adinkras raised with word $w=12$, but merely 
that given a particular valise adinkra, raising with word $w=3$ will lead to the same HYMNs as instead raising the same valise adinkra with word $w=12$. Possibly, 
some adinkras raised with the word $w=3$ may be in the same equivalence class as other 
adinkras raised with the word $w=5$ for instance, it depends on how the valise adinkras are 
related by signed bosons and fermion permutations. 

In order to relate this discussion back to the supermultiplets discussed in chapter three,
let us make a series of observations about the structure of the superfields that describe
them. When {\em {no}} node lowering is applied to the
4D, $\cal N$ = 1 tensor supermultiplet or Hodge Dual \# 3 chiral supermultiplet, under 
the action of projection to 1D, $N$ = 4 supermultiplets, these are within the class of adinkras 
with $w$ = 0.  When {\em {no}} node lowering is applied to the 4D, $\cal N$ = 1 vector 
supermultiplet, the Hodge Dual \# 1, or Hodge Dual \# 2 chiral 4D, $\cal N$ = 1 chiral 
supermultiplets, under the action of projection to 1D, $N$ = 4 supermultiplets, these are 
within the class of adinkras with $w$ = 1,2,4, or 8.  When {\em {no}} node lowering is 
applied to the 4D, $\cal N$ = 1 chiral supermultiplet, under the action of projection to a 
1D, $N$ = 4 supermultiplet, it is within the class of adinkras with $w$ = 3, 5, 6, 9, 10, and 
12.  When {\em {no}} node lowering is applied to the 4D, $\cal N$ = 1 field strengths of 
the tensor supermultiplet, Hodge Dual \# 1, or Hodge Dual \# 2 chiral supermultiplet, under 
the action of projection to 1D, $N$ = 4 supermultiplets, these are within the class of adinkras 
with $w$ = 7, 11, 13, 14.  The spinor field strengths for all of these supermultiplets correspond
to $w$ = 15.
      
\section{Conclusion}
In this paper, we introduced equivalence classes for non-valise adinkras that relate to isomorphisms of adinkras. We defined $B$-matrices and defined HYMNs as the eigenvalues of these matrices. Interestingly, the HYMNs seem to carry all information about isomorphisms in shape. To further the understanding of adinkra isomorphisms of non-valise adinkras, we have developed a program to calculate eigenvalues of $B$-matrices in the Python language that is given in appendix~\ref{a:Python}. A summary of the findings of this program is given in a \emph{Mathematica} notebook 
that can found at the HEPTHools \href{https://hepthools.github.io/Data/}{Data 
Repository} on GitHub.

From Figure~\ref{f:GR22Raised}, it was clear starting from the reference valise adinkra the 
number of other adinkras with lifted nodes that can be constructed from it occur as follows.  
There was one adinkra with no nodes lifted, two adinkras with one node lifted, and finally 
one adinkra with two nodes lifted.  Looking vertically in the column that labels ``Total Number'' 
in Table \ref{Tab:GR22-summary}, we see the ratios of 1:2:1, i.\ e.\ the binomial coefficient 
for two choose an integer.  In turn this implies that the actions of lifting nodes versus using 
flips and flops to generate all sixty-four adinkras commute  one with the other.
       
Looking vertically in the column that labels ``Total Number'' in Table~\ref{Tab:GR44-summary}, 
we see the ratios 1:4:6:4:1 , i.\ e.\ the binomial coefficient for four choose an integer.  As in the 
two color case this also implies that the actions of lifting nodes versus using flips and flops to 
generate all adinkras commute one with the other.

This analysis also provides a simple way to see the number\footnote{This number counts 
both $\chi{}_{\rm o} = \pm 1$ sectors in the work of [24].} of adinkras associated with $BC{}_4$ and 
with all possibility of raised nodes is equal to 4! $\times$ 36,864 = 884,736.

The observations in this work reveal that the HYMNs, i.\ e.\  the eigenvalues of the B-matrices
for the adinkras investigated, can be used to cleanly partition these sets of adinkras 
into distinct classes.  Apparently, the actions of the flipping and flopping operation preserve 
these classes.  This offers an explicit route whereby considering instead of individual adinkras 
as the basis for higher dimensional supermultiplets, it is the emergent class structure of adinkras 
that provides such a basis.

The introduction of the matrix $M(m,w)$ as the nodal raising operator has implications 
for future directions of research.  Previous mathematical ``devices'' introduced for the
analysis of adinkras include holoraumy \cite{HL1} and the Gadget \cite{Gd1}.  But all
such previous discussions have been restricted to valise adinkras.  Clearly these can
now be modified by appropriate introductions of $M(m,w)$ into their definitions.  As
well, future study of the holoraumy, Gadget, HYMNs, and $\chi{}_{\rm o}$ all seems 
indicated to ascertain their dependence on adinkra dashing as well as the impacts
of $M(m,w)$.  Another class of questions to study is the generalization of this formalism 
to the cases where valise nodes can be lifted more than once.  The simplest place to 
study this is for three-color adinkras. Finally with the introduction of $M(m,w)$ a 
re-examination of the work in \cite{EH1} and \cite{EH2} is also possible to be extended 
to the entire 36,864 $BC{}_4$ related adinkras.

  \vspace{.05in}
 \begin{center}
\parbox{4in}{{\it ``Don't lower your expectations to meet your performance. Raise your 
level of performance to meet your expectations.'' \\ ${~}$ 
\\ ${~}$ }\,\,-\,\, Ralph Marston $~~~~~~~~~$}
 \parbox{4in}{
 $~~$}  
 \end{center}

\noindent
{\bf Added Note In Proof}\\[.1in] \indent
After the completion of this work, the relevance of the research works in \cite{Top1,Top2,Top3,Top4}
were brought to our attention.  The interested reader is referred to these papers to see the
relation of ``dressing matrices'' (introduced in these works) and our height-raising factors.

\noindent
{\bf Acknowledgments}\\[.1in] \indent
This research is supported in part by the endowment of the Ford Foundation Professorship 
of Physics at Brown University. Yangrui Hu would like to acknowledge her participation 
in the second annual Brown University Adinkra Math/Phys Hangout" during 2017. This 
work was partially supported by the National Science Foundation grant PHY-1620074. 
S.\ J.\ G.'s research in the work was also supported in part by the endowment of the 
John S.~Toll Professorship, the University of Maryland Center for String \& Particle 
Theory, National Science Foundation Grant PHY-09-68854.  S.\ J.\ G. acknowledges 
the generous support of the Roth Professorship and the very congenial and generous 
hospitality of the Dartmouth College physics department in the period of this investigation.
Finally, we would like to acknowledge Mr.\ Isaac Friend for his cross-checking a previous
version of the code used to sort the BC${}_4$ related adinkras.

\newpage
\appendix
\section{Generating Minimal 1D SUSY Representations}\label{a:Recursion}

\label{sec:recurrence}
$~~~~$ When the results for ${{\rm d}}{}_{min}(N)$ (described in the introduction) are written
more explicitly for values 1 $\le$ $N$ $\le$ 16, they can be expressed
as shown in the first table.
\begin{table}[h]
\caption{ Number of Supercharges vs. Number of Bosons (or Fermions) \label{tab:I}}
\vspace{0.2cm}
\begin{center}
\footnotesize
\begin{tabular}{|c||c|c|c|c|c|c|c|c|c|c|c|c|c|c|c|c|}
\hline
$N$  &1&2&3&4&5&6&7&8&9&10&11&12&13&14&15&16\\\hline
$\rdm(N)$&1&2&4&4&8&8&8&8&16&32&64&64&128&128&128&128\\\hline
\end{tabular}
\end{center}
\label{table:dmin}
\end{table} \vskip0.01pt \noindent
From this table we see there are some values of $N$ such that as one goes from 
$N$ to $N+1$, the size of the L-matrices and R-matrices ``jumps'' by a factor of two.  
This occurs for all the $N$ values in the sequence given by
\be
    1 , ~ 2, ~ 4, ~ 8, ~ 9, ~ 10, ~12, ~ 16, ~17, ~18, ~20, ~25, \dots
\ee
and we can borrow language from nuclear physics\footnote{See the webpage 
at https://en.wikipedia.org/wiki/Magic$_-$number$_-$(physics) on-line.} 
and call these ``magical'' values of $N$.  

This chart also illustrates the equation
\be
\rdm(N \,+\, 8) ~=~ 16 \, \rdm (N)   ~~~,
\label{dm1}
\ee
or more generally (going beyond the values of the table),
\be
\rdm(N \,+\, 8 m ) ~=~ 16^{m} \, \rdm(N)  ~~~.
\label{dm2}
\ee
This implies that the magical numbers are periodic with period $8$.  This is reminiscent 
of Bott periodicity, where the homotopy groups of the infinite dimensional real orthogonal 
group $O$ is given by a periodic sequence of groups with periodicity $n$ (mod 8): $\pi_k
(O) = \pi_{k+ 8} (O)$\cite{Bott}.  Here, the magic numbers are one more than 
the degrees in which these groups are non-trivial.  The connection between these phenomena 
is due to the relationship of  ${\cal {GR}}$(d,$N)$ algebras to Clifford algebras, with the 
latter also exhibiting the period $8$ behavior \cite{rLM}.

In the following, we will describe the recursive algorithm introduced in \cite{GRana1,GRana2} 
as well as bring our results in line with our current conventions and notation.  This will 
also provide justifications for the formulae in (\ref{dm1}) and (\ref{dm2}).

As will be seen in each case when the value of $N$ is ``magical," one $L$-matrix can be 
chosen to be equal to the corresponding $R$-matrix and both are symmetric under matrix 
transposition.  In fact, this one matrix is taken to be the identity matrix.  The remaining $(N-1)$ 
$L$-matrices are equal to the negative of the corresponding $R$-matrices and both are 
antisymmetric under matrix transposition.  The recursion construction only makes use of the 
antisymmetrical matrices that occur in the magical values of $N$.

For $N = 2$, there were given the set of $2 \times 2$ matrices: 
\be\eqalign{
{  L }_{1} ~&=~  {  {\rm I}}{}_{2 \times 2} ~=~~~~~\, R_{1}  ~~~ ~,~~  \cr
 L _{2} ~&=~ i{  \s}^2 ~~=~ - \, R_{2}
 ~~~~ . } \label{N2} \ee 

\noindent For $N = 4$, there were given two distinct minimal set of matrices that realize 
the $\GR(4, 4)$ algebra given by
\be \begin{array}{ccccccccc}
L _{1} &=&  \, {  {\rm I}}{}_{2 \times 2} & \otimes & {  {\rm I}}{}_{2 
\times 2} &=& ~~&R_{1} & ~~~, \\
L _{2} &=& i{  \s}^1 & \otimes &{  \s}^2 &=& - &R_{2}&  ~~~, \\
L _{3} &=& i{  \s}^2 & \otimes & {  {\rm I}}{}_{2 \times 2} &=& - &{  
{\rm R}}_{3}&  ~~~, \\
L _{4} &=& -i{  \s}^3 & \otimes &{  \s}^2 &=& - &R_{4}&  ~~~, \\ 
 & &  &  & & & & & \\ 
{\Tilde L }_{1} &=&  \, {  {\rm I}}{}_{2 \times 2} & \otimes & {  {\rm I}}{}_{2 \times 2} 
 &=&  ~ &{\Tilde  R}_{1}&  ~~~,  \\ 
{\Tilde L }_2 &=& i{  \s}^2 & \otimes  &{  \s}^3 &=& 
- &{\Tilde  R}_2&  ~~~, \\ 
{{\Tilde L} }_3 &=& - i {  {\rm I}}{}_{2 \times 2} & \otimes  &{  \s}^2 &=& - 
&{\Tilde  R}_3&  ~~~, \\ 
{\Tilde L }_4 &=& i{  \s}^2  & \otimes &{  \s}^1 &=& -
&{\Tilde R}_4&   ~~~. \\ 
\end{array} \label{N4}  \ee
Any three within each of these given sets can be used to cover the case of $N$ = 3.
\noindent For $N = 8$, a convenient set for our required matrices is given by,
\be  \begin{array}{cccccccccccc}
L _{1} &=& \,  {  {\rm I}}{}_{2 \times 2}\ & \otimes & {  {\rm I}}{}_{2 \times 
2} & \otimes & {  {\rm I}}{}_{2 \times 2}  &=& &R_{1}& &~~~, ~~~ \\
L _{2} &=& i {  {\rm I}}{}_{2 \times 2} & \otimes &{  \s}^3 & \otimes &{  
\s}^2 &=& - &R_{2}& &~~~, ~~~ \\
L _{3} &=& i{  \s}^3 & \otimes &{  \s}^2 & \otimes & {  {\rm I}}{}_{2 
\times 2} &=& -  &R_{3}& &~~~, ~~~ \\
L _{4} &=& i {  {\rm I}}{}_{2 \times 2} & \otimes &{  \s}^1 & \otimes &
{  \s}^2 &=& - &R_{4}& &~~~, ~~~ \\
L _{5} &=& i{  \s}^1 & \otimes &{  \s}^2 & \otimes & {  {\rm I}}{
}_{2 \times 2} &=& - &R_{5}& &~~~, ~~~ \\
 L _{6} &=& i{  \s}^2 & \otimes & {  {\rm I}}{}_{2 \times 2}  & \otimes 
 & {  \s}^1&=& -  &R_{6}& &~~~, ~~~ \\
L _{7} &=& i{  \s}^2 & \otimes & {  {\rm I}}{}_{2 \times 2} & \otimes &
{  \s}^3 &=& - &R_{7}&  &~~~, ~~~ \\
L _{8} &=& i{  \s}^2 & \otimes &{  \s}^2 & \otimes &{  \s}^2 
&=& -  &R_{8}& &~~~. ~~~ \\
\end{array}
\label{N8} \ee
For the cases of $N = 5$, $6$, and $7$, these can be formed by taking any subset consisting 
of $5$, $6$, or $7$ elements of the set of $N$ = 8 matrices, respectively. 

Finally, in the works of \cite{GRana1,GRana2} the existence of a recursion formula\footnote{The 
formulae which appear in (\ref{ReCuRs}) make corrections with regard to the modification 
$n \times n \to \rdm(n)  \times \rdm(n)$.} generating matrix representations for arbitrary values 
of $m$ and $n$ in the formula $N = 8m + n$ (where $m$ is any non-negative integer and $n$ 
is an integer so that $1 \le n \le 8$) was given. We define,
\be  \begin{array} {cccccccccccc}
L _{1} &=& {  {\rm I}}{}_{2 \times 2} & \otimes & {  {\rm I}}{}_{\rdm(n) \times \rdm(n)}  
& \otimes & {  {\rm I}}{}_{\rdm(8m) \times \rdm(8m)} &=&  &R_{1}& 
&~~~,\\
L _{2} &=& i {  \s}^2 & \otimes & {  {\rm I}}{}_{\rdm(n)  \times \rdm(n)}  & \otimes 
& {  {\rm I}}{}_{\rdm(8m) \times \rdm(8m)}  &=&  - &R_{2}& 
&~~~,  \\
L _{\widehat {\rm A}} &=& {  \s}^3 &\otimes & L _{\widehat {\rm A}}
(n)_{\rdm(n)  \times \rdm(n)}  & \otimes & {  {\rm I}}{}_{\rdm(8m) \times \rdm(8m)}  &=&  -  
&R_{\widehat {\rm A}}& & ~~~,\\
L _{\widehat {\rm M}} &=& {  \s}^1 & \otimes & {  {\rm I}}{}_{\rdm(n)  
\times \rdm(n)} & \otimes & L _{\widehat {\rm M}}(8m)_{\rdm(8m) \times \rdm(8m)} &=&  -  
&{  {\rm R
}}_{\widehat {\rm M}}& &~~~,  \\
\end{array}
\label{ReCuRs} \ee \vskip1pt \noindent
and in the following we include discussion on why this works.  However, we note there 
exist many ways to construct such recursion formulae.  One source of this diversity in 
the fact that the roles of the ${  \s}^1$ and ${  \s}^3$ matrices can be ``swapped.''

Above, we have used the notation where ${  {\rm I}}{}_{\rdm(n) \times \rdm(n) } $ represents 
the $\rdm(n)  \times \rdm(n)$ identity matrix for a given $n$ where $\rdm(n)$ can be read from 
the function in (\ref{e:dmin}) by putting $m=0$.  The matrices $L _{\widehat {\rm A}}(n)_{
\rdm(n)  \times \rdm(n) }$ and $L _{\widehat {\rm M}}(8m)_{\rdm(8m) \times \rdm(8m)} $ are 
only taken from the purely antisymmetrical L-matrices for any value of $N$.   Under this 
circumstance the index ${\widehat {\rm A}}$ is restricted and chosen to only run over those 
${  {\rm L}}_{\widehat {\rm A}}(n)$ matrices that are antisymmetric.  The index ${\widehat {
\rm M}}$ is restricted and chosen to only run over those $L _{\widehat {\rm M}}(8m)_{\rdm(
8m)\times \rdm(8m)}$ matrices that are antisymmetric matrices for the case of $N$ = $8m$. 
Application of this recursion formulae to the previous cases lead to the results reported in 
the following discussion.  It is instructive to spend a bit of time on the topic of results that 
arise from the recursion formula in (\ref{ReCuRs}).


Clearly $9 = 8 + 1$ which implies for this case we should use the recursion formula with 
$m = 1$ and $n = 1$.  For $n = 1$, $L _1=R_1=1$.  In particular this means that the 
second factor is absent.  But also there are no $L _{\widehat {\rm A}}(1)_{1 \times 1}$ 
matrices.  In  this case, the recursion formula (\ref{ReCuRs}) collapses to 
\be  \begin{array} {cccccccccccc}
L _{1} &=& {  {\rm I}}{}_{2 \times 2} & \otimes & {  {\rm I}}{}_{8
\times 8}  & = & ~~R_{1} &~,&  &~&  &~~~\\
L _{\widehat {\rm A}} &=& i {  \s}^2 &\otimes & 
{  {\rm I}}{}_{8 \times 8}   &= & - \, R_{\widehat {\rm A}} &~~\,,~&  ~  & ~& & ~~~\\
L _{\widehat {\rm M}} &=& {  \s}^1 & \otimes & 
L _{\widehat {\rm M}}(8) 
& = & - \, R_{\widehat {\rm M}}  &\, \, ,&    &~& &~~~  \\
\end{array}
\label{ReCuRs2} \ee \vskip1pt \noindent 
and the  $L _{\widehat {\rm M}}(8) $
matrices are simply the last seven matrices as these are the anti-symmetrical ones
in (\ref{N8}) for $N$ = 8.  Thus, for $N$ = 9, we find the explicit form of the 16 $\times$ 16 
matrices are given by:
\be \begin{array}{ccccccccccccc}
L {}_1&=& {  {\rm I}}{}_{2 \times 2}& \otimes & {  {\rm I}}{}_{2 \times 2} & \otimes & 
{  {\rm I}}{}_{2 \times 2}&\otimes & {  {\rm I}}{}_{2 \times 2}&=& & R{}_1&  ~~~, \\
 L {}_2&=&  i {  \s}^2  & \otimes & {  {\rm I}}{}_{2 \times 2}
 & \otimes &  {  {\rm I}}{}_{2 \times 2} &  \otimes &
 {  {\rm I}}{}_{2 \times 2} & =& -  &R{}_2& ~~~,  \\
L {}_3&=& i{  \s}^1 & \otimes & {  {\rm I}}{}_{2 \times 2} & \otimes &
{  \s}^3  &  \otimes &  {  \s}^2 & =& -  &R{}_3&  ~~~,  \\
L {}_4 &=& i{  \s}^1 & \otimes &{  \s}^3 & \otimes & {  \s}^2 &
 \otimes & {  {\rm I}}{}_{2 \times 2}
 & =& -  &R{}_4&  ~~~,  \\
L {}_5&=& i{  \s}^1 & \otimes & {  {\rm I}}{}_{2 \times 2} & \otimes & {  \s}^1 &  
\otimes & {  \s}^2 &=& -  &R{}_5& ~~~, \\
L {}_6&=& i{  \s}^1& \otimes & {  \s}^1 & \otimes & {  \s}^2 & 
\otimes & {  {\rm I}}{}_{2 \times 2} & =& -  &R{}_6&  ~~~,  \\
L {}_7 &=& i{  \s}^1 & \otimes &{  \s}^2 & \otimes &
{  {\rm I}}{}_{2 \times 2} & \otimes & {  \s}^1& =& -  &R{}_7&  ~~~, \\
L {}_8&=& i{  \s}^1 & \otimes &{  \s}^2 & \otimes &
{  {\rm I}}{}_{2 \times 2} & \otimes & {  \s}^3& =& -  &R{}_8&  ~~~, \\
L {}_9&=& i{  \s}^1 & \otimes & {  \s}^2 & 
\otimes & {  \s}^2 & \otimes &{  \s}^2
& =& -  &R{}_9&  ~~~.
\end{array} \label{N9a}  \ee

Since 10 = 8 + 2, this implies for this case we have $m$ = 1 and $n$ = 2.  The L-matrices
and R-matrices for $n$ = 2 appear in (\ref{N2}).  Thus, we have ${  {\rm I}}{}_{\rdm(n)  
\times \rdm(n) }$ = ${  {\rm I}}{}_{2 \times 2}$ and the set $  L _{\widehat {\rm A}}({2 
\times 2})$ consists of the single element $i {  \s}^2$.  Using these, the recursion formula 
yields
\be  \begin{array} {cccccccccccc}
L _{1} &=& {  {\rm I}}{}_{2 \times 2} & \otimes & {  {\rm I}}{}_{2
\times 2}  & \otimes & {  {\rm I}}{}_{8 \times 8} &=&  &R_{1}& 
&~~~;\\
L _{2} &=& i {  \s}^2 & \otimes & {  {\rm I}}{}_{2 \times 2}  & \otimes 
& {  {\rm I}}{}_{8 \times 8}  &=&  - &R_{2}& 
&~~~;  \\
L _{3} &=& i {  \s}^3 &\otimes &  {  \s}^2 
& \otimes & {  {\rm I}}{}_{8 \times 8}  &=&  -  &R_{3}& & ~~~; \\
L _{\widehat {\rm M}} &=& {  \s}^1 & \otimes & {  {\rm I}}{}_{2 \times 2} 
& \otimes & L _{\widehat {\rm M}}(8)  &=&  -  &{  {\rm R
}}_{\widehat {\rm M}}& &~~~;  \\
\end{array}
\label{N10z} \ee \vskip1pt \noindent
where $L _{\widehat {\rm M}}(8)$ once again consists of the seven antisymmetrical 
matrices that occur in the case of ${8 \times 8}$ matrices that are explicitly exhibited
in (\ref{N8}).

As will be described in the equation (\ref{ReCuRs}), the recursion formula generates 
successively larger and larger matrices that satisfy a set of algebraic conditions.  These 
conditions are those required to realize $N$ supercharges linearly on $d_{\rm {min}}$
bosons and the same number of fermions.  It also plays the very important role of extending 
the matrix representation that have $N$ $\le$ 8 to those which have  $N$ $>$ 8.  The 
recursion formula is also the basis for the answer to the first question raised in this work 
as well as the origin of the ``magic number'' sequence.

More explicitly for $N$ = 10 we able to find a $32 \times 32$ representation:
\be
\begin{array}{cccccccccccccccc} 
L _{1} &=&{  {\rm I}}{}_{2 \times 2} &\otimes & {  {\rm I}}{}_{2 \times 2} 
& \otimes & {  {\rm I}}{}_{2 \times 2}& \otimes & {  {\rm I}}{}_{2 \times 2} & \otimes 
&{  {\rm I}}{}_{2 \times 2} & =&  &R_{1}& , \\
L _{2} &=& i{  \s}^2 &\otimes & {  {\rm I}}{}_{2 \times 2} & \otimes & 
{  {\rm I}}{}_{2 \times 2}& \otimes & {  {\rm I}}{}_{2 \times 2} & \otimes &{  
{\rm I}}{}_{2 \times 2} & =&- &R_{2}& , \\
L _{3} &=& i{  \s}^3 &\otimes &{  \s}^2 &\otimes& {  {\rm I}}{}_{
2 \times 2}& \otimes &{  {\rm I}}{}_{2 \times 2} &\otimes & {  {\rm I}}{}_{2 
\times 2}& =&-  &R_{3}&,   \\
L _{4}  &=&i  {  \s}^1 &\otimes & {  {\rm I}}{}_{2 \times 2} &\otimes &
{  {\rm I}}{}_{2 \times 2} & \otimes &{  \s}^3 & \otimes &{  \s}^2
&=&-  &R_{4}&,  \\
L _{5} &=& i {  \s}^1& \otimes & {  {\rm I}}{}_{2 \times 2} &\otimes &
{  \s}^3 & \otimes &{  \s}^2 & \otimes & {  {\rm I}}{}_{2 \times 2} 
&=&-  &R_{5}&,  \\
L _{6} &=&i {  \s}^1 &\otimes & {  {\rm I}}{}_{2 \times 2} & \otimes &
{  {\rm I}}{}_{2 \times 2} & \otimes &{  \s}^1 & \otimes &{  \s}^2 
& =&-  &R_{6}& , \\
L _{7} &=& i {  \s}^1 & \otimes & {  {\rm I}}{}_{2 \times 2} & \otimes &
{  \s}^1 & \otimes &{  \s}^2 & \otimes & {  {\rm I}}{}_{2 \times 2} 
& =&-  &R_{7}& , \\
L _{8} &=& i {  \s}^1&\otimes & {  {\rm I}}{}_{2 \times 2} & \otimes & 
{  \s}^2 & \otimes & {  {\rm I}}{}_{2 \times 2}  & \otimes & {  \s}^1
& =&-  &R_{8}& , \\
L _{9} &=& i {  \s}^1 &\otimes & {  {\rm I}}{}_{2 \times 2} & \otimes &
{  \s}^2 & \otimes & {  {\rm I}}{}_{2 \times 2} & \otimes &{  \s}^3 
& =&-  &R_{9}& , \\
L _{10} &=& i {  \s}^1 &\otimes & {  {\rm I}}{}_{2 \times 2} & \otimes & 
{  \s}^2 & \otimes &{  \s}^2 & \otimes &{  \s}^2
& =& - &\,R_{10}& . \\
 \end{array} \label{N10} \ee
\indent The next magical value is the $N$ = 12 case to lead to a 64 $\times$ 64 representation and
12 = 8 + 4, implies for this case we have $m$ = 1 and $n$ = 4.  
Thus, we have ${  {\rm I}}{}_{\rdm(n)  \times \rdm(n) }$
= ${  {\rm I}}{}_{4 \times 4}$, the $L _{\widehat {\rm A}}({4 \times 4})$ 
set consists of the final three elements from the matrices that appear in (\ref{N4}), and the 
set $L _{\widehat {\rm M}}({8 \times 8})$ 
set consists of the final seven elements from the matrices that appear in (\ref{N8}).
\be  \begin{array} {cccccccccccc}
L _{1} &=& {  {\rm I}}{}_{2 \times 2} & \otimes & {  {\rm I}}{}_{
4 \times 4}  & \otimes & {  {\rm I}}{}_{8 \times 8} &=&  &R_{1}& 
&~~~,\\
L _{2} &=& i {  \s}^2 & \otimes & {  {\rm I}}{}_{4 \times 4}  & \otimes 
& {  {\rm I}}{}_{8 \times 8}  &=&  - &R_{2}& 
&~~~,  \\
L _{p} &=& {  \s}^3 &\otimes & L _{p}
(4)  & \otimes & {  {\rm I}}{}_{8 \times 8}  &=&  -  &
R_{p}& & ~~~,\\
L _{q} &=& {  \s}^1 & \otimes & {  {\rm I}}{}_{4 \times 4} 
& \otimes & L _{q}(8)  &=&  -  &{  {\rm R
}}_{q}& &~~~,  \\
\end{array}
\label{ReCuRsVxy} \ee 
\noindent
where we have used the compact notation to efficiently express the forms of the
appropriate $64 \times 64$ matrices.  In this expression, the index $p$ takes on
the values of $p$ = 3, 4, and 5 while the index $q = 6, \dots, 12$.
More explicitly this becomes
\be
\begin{array}{ccccccccccccccc}
L _{1}&=& {  {\rm I}}{}_{2 \times 2} & \otimes &{  {\rm I}}{}_{2 \times 2}& \otimes&{  {\rm I}}
{}_{2 \times 2}& \otimes &{  {\rm I}}{}_{2 \times 2} & \otimes & {  {\rm I}}{}_{2 \times 2} & \otimes & {  
{\rm I}}{}_{2 \times 2} & = &  R_{1} ~~~  ,\\
L _2& = &i {  {\s}}^2 & \otimes &{  {\rm I}}{}_{2 \times 2}& \otimes & {  {\rm I}}{}_{2 \times 2} 
& \otimes &{  {\rm I}}{}_{2 \times 2} & \otimes & {  {\rm I}}{}_{2 \times 2} & \otimes & {  {\rm I}}{}_{2 
\times 2} & = & \,-\, R_2 ~~ ,~~\\ 
L _3 & = & i {  {\s}}^3 & \otimes & {  {\s}}^1 & \otimes & {  {\s}}^2 & \otimes & {  {\rm I}}{}_{
2 \times 2} & \otimes & {  {\rm I}}{}_{2 \times 2} & \otimes & {  {\rm I}}{}_{2 \times 2} & = & -\, R_3
~~~ , ~~\\ 
L _4 & = & i {  {\s}}^3 & \otimes & {  {\s}}^2 & \otimes & {  {\rm I}}{}_{2 \times 2} & \otimes & 
{  {\rm I}}{}_{2 \times 2} & \otimes  & {  {\rm I}}{}_{2 \times 2} & \otimes & {  {\rm I}}{}_{2 \times 2} & = 
& - \,R_4 ~~~ , ~~\\ 
L _5&=&i {  {\s}}^3 & \otimes &{  {\s}}^3& \otimes&{  {\s}}^2 &\otimes &{  {\rm I}}{}_{2 \times 
2} & \otimes & {  {\rm I}}{}_{2 \times 2} & \otimes & {  {\rm I}}{}_{2 \times 2} & = & - \, R_5 
~~~ , ~~\\ 
L _{6}&=&i {  {\s}}^1 & \otimes &{  {\rm I}}{}_{2 \times 2}& \otimes&
{  {\rm I}}{}_{2 \times 2}& \otimes &  {  {\rm I}}{}_{2 \times 2}
 & \otimes & {  {\s}}^3 & \otimes & {  {\s}}^2 & = & -\,  R_{6} ~~~ , ~~\\ 
L _7&=&i {  {\s}}^1 & \otimes &{  {\rm I}}{}_{2 \times 2}& \otimes&
{  {\rm I}}{}_{2 \times 2} &\otimes &{  
{\s}}^3 & \otimes &  {  {\s}}^2  & \otimes & {  {\rm I}}{}_{2 \times 2} & = & - \, R_7  
~~~ , ~~\\ 
L _8&=&i {  {\s}}^1 & \otimes &{  {\rm I}}{}_{2 \times 2}& \otimes &{  {\rm I}}{}_{2 \times 2} &
\otimes & {  {\rm I}}{}_{2 \times 2} & \otimes & {  {\s}}^1 & \otimes &  {  {\s}}^2 
 & = & - \, R_8 
~~~ , ~~\\ 
L _9&=&i {  {\s}}^1 & \otimes &{  {\rm I}}{}_{2 \times 2}& \otimes&{  {\rm I}}{}_{2 \times 2}& 
\otimes &{  {\s}}^2 & \otimes & {  {\s}}^3 & \otimes & {  {\rm I}}{}_{2 \times 2} & = & - \, R_9 
~~~ , ~~\\ 
L _{10}&=&i {  {\s}}^1 & \otimes &{  {\rm I}}{}_{2 \times 2}& \otimes&
{  {\rm I}}{}_{2 \times 2} 
& \otimes & {  {\s}}^2 & \otimes &
{  {\rm I}}{}_{2 \times 2} 
& \otimes &
 {  {\s}}^1 & = & - \, R_{10} ~~~ , ~~\\ 
L _{11}&=&i {  {\s}}^1 & \otimes &{  {\rm I}}{}_{2 \times 2}& \otimes&
{  {\rm I}}{}_{2 \times 2} 
& \otimes & {  {\s}}^2 & \otimes &
{  {\rm I}}{}_{2 \times 2} 
& \otimes &
 {  {\s}}^3 & = & - \, R_{11} ~~~ , ~~\\ 
L _{12}&=&i {  {\s}}^1 & \otimes &{  {\rm I}}{}_{2 \times 2}& \otimes&
{  {\rm I}}{}_{2 \times 2}& \otimes &{  
{\s}}^2 & \otimes & {  {\s}}^2 & \otimes & {  {\s}}^2 & = & -\,  R_{12} ~~~ . ~~\\ 
\end{array}
\label{N12} 
\ee
The case of $N$ = 11 is contained as a subset. 

The final explicit result that we present is for $N$ = 16, where 
we have a 128 x 128 representation of the $N$ = 16 supersymmetry algebra:
\be  \begin{array} {cccccccccccc}
L _{1} &=& {  {\rm I}}{}_{2 \times 2} & \otimes & {  {\rm I}}{}_{
8 \times 8}  & \otimes & {  {\rm I}}{}_{8 \times 8} &=&  &R_{1}& 
&~~~,\\
L _{2} &=& i {  \s}^2 & \otimes & {  {\rm I}}{}_{8 \times 8}  & \otimes 
& {  {\rm I}}{}_{8 \times 8}  &=&  - &R_{2}& 
&~~~,  \\
L _{r} &=& {  \s}^3 &\otimes & L _{r}
(8)  & \otimes & {  {\rm I}}{}_{8 \times 8}  &=&  -  &
R_{r}& & ~~~,\\
L _{s} &=& {  \s}^1 & \otimes & {  {\rm I}}{}_{8 \times 8} 
& \otimes & L _{s}(8)  &=&  -  &{  {\rm R
}}_{s}& &~~~,  \\
\end{array}
\label{ReCuRsWp} \ee 
\noindent
where we have again used a compact notation to efficiently express the forms of the
appropriate $128 \times 128$ matrices.  In this expression, the index $r$ takes on
the values of $r$ = 3, $\dots$, 9 while the index $s$ = 10, $\dots$, 16. Expanding
this out completely yields,
\be
\begin{array}{ccccccccccccccccrc}
L _{1} &=& {  {\rm I}}{}_{2 \times 2} & \otimes & {  {\rm I}}{}_{2 \times 2} & \otimes & {  {\rm I}}{}_{2 
\times 2} & \otimes & {  {\rm I}}{}_{2 \times 2}&\otimes& {  {\rm I}}{}_{2 \times 2}&\otimes &{  {\rm I}}{}_{2 
\times 2} &\otimes &{  {\rm I}}{}_{2 \times 2} &=&  R_{1} &, \\
L _2 & =& i {  {\s}}^2 &\otimes & {  {\rm I}}{}_{2 \times 2}& \otimes & {  {\rm I}}{}_{2 \times 2} & \otimes 
& {  {\rm I}}{}_{2 \times 2} & \otimes & {  {\rm I}}{}_{2 \times 2} & \otimes & {  {\rm I}}{}_{2 \times 2} & \otimes & 
{  {\rm I}}{}_{2 \times 2} &  = &  - \, R_2 &, \\
L _3 & =& i {  {\s}}^3 &\otimes & {  {\rm I}}{}_{2 \times 2} & \otimes &  {  {\s}}^3 & \otimes & {  {\s}}^2 &
 \otimes& {  {\rm I}}{}_{2 \times 2} & \otimes & {  {\rm I}}{}_{2 \times 2} & 
\otimes &{  {\rm I}}{}_{2 \times 2} &=&   - \, R_3 &, \\
L _4 & =& i {  {\s}}^3&\otimes& {  {\s}}^3 & \otimes& {  {\s}}^2 & \otimes & {  {\rm I}}{}_{2 \times 2}
&\otimes & {  {\rm I}}{}_{2 \times 2} & \otimes &{  {\rm I}}{}_{2 \times 2}& \otimes &{  {\rm I}}{}_{2 \times 2} &=&   
- \, R_4 &, \\
L _5 &=&i {  {\s}}^3 &\otimes &{  {\rm I}}{}_{2 \times 2}& \otimes &{  {\s}}^1 &\otimes &{  {\s}}^2 &\otimes &
{  {\rm I}}{}_{2 \times 2}& \otimes & {  {\rm I}}{}_{2 \times 2} & \otimes & {  {\rm I}}{}_{2 \times 2} &=& - \, R_5 &, \\
L _6 &=& i {  {\s}}^3 &\otimes &{  {\s}}^1 &\otimes& {  {\s}}^2 & \otimes &
 {  {\rm I}}{}_{2 \times 2} &\otimes 
&{  {\rm I}}{}_{2 \times 2}&\otimes & {  {\rm I}}{}_{2 \times 2}&\otimes &{  {\rm I}}{}_{2 \times 2}&=& - \, R_6 &, \\
L _7 &=&i {  {\s}}^3 &\otimes &{  {\s}}^2 &\otimes& {  {\rm I}}{}_{2 \times 2}& \otimes &{  {\s}}^1 &\otimes
&{  {\rm I}}{}_{2 \times 2}&\otimes& {  {\rm I}}{}_{2 \times 2} & \otimes &{  {\rm I}}{}_{2 \times 2}&=& - \, R_7 &, \\
L _8 &=&i {  {\s}}^3 &\otimes &{  {\s}}^2 &\otimes& {  {\rm I}}{}_{2 \times 2}& \otimes &{  {\s}}^3 &\otimes
&{  {\rm I}}{}_{2 \times 2}&\otimes& {  {\rm I}}{}_{2 \times 2} & \otimes &{  {\rm I}}{}_{2 \times 2}&=& - \, R_8 &, \\
L _9 &=&i{  {\s}}^3&\otimes&{  {\s}}^2 &\otimes &{  {\s}}^2 &\otimes &{  {\s}}^2 &\otimes & {  {\rm I}}{}_{2 \times 2} 
&\otimes& {  {\rm I}}{}_{2 \times 2}&\otimes &{  {\rm I}}{}_{2 \times 2} &=&- \, R_9&, \\
L _{10} &=& i {  {\s}}^1& \otimes & {  {\rm I}}{}_{2 \times 2}& \otimes& {  {\rm I}}{}_{2 \times 2}& \otimes 
&{  {\rm I}}{}_{2 \times 2}&\otimes&{  {\rm I}}{}_{2 \times 2}&\otimes&{  {\s}}^3&\otimes &{  {\s}}^2&=&- \, R_{10}&,\\
L _{11} &=&i {  {\s}}^1 &\otimes&{  {\rm I}}{}_{2 \times 2}&\otimes&{  {\rm I}}{}_{2 \times 2}& \otimes 
&{  {\rm I}}{}_{2 \times 2}&\otimes &{  {\s}}^3 &\otimes &{  {\s}}^2 &\otimes&{  {\rm I}}{}_{2 \times 2}&=&- \, R_{11}&,\\
L _{12} &=& i {  {\s}}^1 &\otimes&{  {\rm I}}{}_{2 \times 2}&\otimes&{  {\rm I}}{}_{2 \times 2}&\otimes 
&{  {\rm I}}{}_{2 \times 2}&\otimes&{  {\rm I}}{}_{2 \times 2}&\otimes&{  {\s}}^1&\otimes&{  {\s}}^2&=&- \, R_{12}  &, \\
L _{13} &=&i {  {\s}}^1 &\otimes& {  {\rm I}}{}_{2 \times 2}& \otimes &{  {\rm I}}{}_{2 \times 2}&\otimes 
&{  {\rm I}}{}_{2 \times 2}& \otimes &{  {\s}}^1 &\otimes& {  {\s}}^2 &\otimes& {  {\rm I}}{}_{2 \times 2}&=&- \, R_{13}&,\\
L _{14} &=& i {  {\s}}^1 &\otimes& {  {\rm I}}{}_{2 \times 2}& \otimes& {  {\rm I}}{}_{2 \times 2}&\otimes 
&{  {\rm I}}{}_{2 \times 2}&\otimes &{  {\s}}^2 &\otimes&{  {\rm I}}{}_{2 \times 2}& \otimes &{  {\s}}^1 &=& - \, R_{14}&,\\
L _{15} &=& i {  {\s}}^1 &\otimes& {  {\rm I}}{}_{2 \times 2}&\otimes& {  {\rm I}}{}_{2 \times 2}&\otimes 
&{  {\rm I}}{}_{2 \times 2}& \otimes &{  {\s}}^2 &\otimes& {  {\rm I}}{}_{2 \times 2}&\otimes &{  {\s}}^3 &=&- \, R_{15}&,\\
L _{16} &=&i {  {\s}}^1 &\otimes&{  {\rm I}}{}_{2 \times 2}&\otimes&{  {\rm I}}{}_{2 \times 2}&\otimes
&{  {\rm I}}{}_{2 \times 2} &\otimes &{  {\s}}^2& \otimes &{  {\s}}^2& \otimes &{  {\s}}^2 &=&- \, R_{16} &.\\
\end{array}
\label{N16} \ee 
\noindent
The cases of $N$ = 13, 14, and 15 are contained as subsets.

A careful comparison between all the results above and those presented in the 
works of \cite{GRana1,GRana2} shows they are not identical.  While both sets
of results are correct, what we have done in the present work is to {\em {actually}}
use the recursion formula to generate the cases of $9 \le N \le 16$.

\newpage
\section{Original script for the program in the Python language}\label{a:Python}

\perlscript{Analysis_GR44}{Original script for the program in the Python language\label{script}}
  
\newpage

\end{document}